\documentclass{aa}

\usepackage{epsfig,graphicx,natbib}
\usepackage{longtable}
\usepackage{amssymb}
\usepackage{amsmath}
\usepackage{mathrsfs} 
\usepackage{color}
\usepackage{subcaption}
\usepackage{booktabs}
\usepackage{xspace}
\usepackage{float}
\usepackage{longtable}
\usepackage{breqn}
\usepackage{algpseudocode}
\usepackage{algorithm}
\usepackage{hyperref}

\usepackage{todonotes}

\newcommand{\norm}[1]{\lVert #1 \rVert}

\newcommand{\RR}{\mathbb{R}}

\newcommand{\pfront}{P_{\RR^m}\left(F\right)}

\begin{document}

\title{Using multiobjective optimization to reconstruct interferometric data (I)}

\author{Hendrik M\"uller \inst{1}
\and Alejandro Mus \inst{2,3}\thanks{Both first authors have contributed equally to this work.}
\and Andrei Lobanov \inst{1}
%\and Gordon Shumway \inst{4}
}

\institute{
  Max-Planck-Institut für Radioastronomie, Auf dem Hügel 69, D-53121 Bonn (Endenich), Germany \\ \email{hmueller@mpifr-bonn.mpg.de}
  \and
  Departament d’Astronomia i Astrof\'isica, Universitat de Val\`encia, C. Dr. Moliner 50, 46100 Burjassot ,Val\`encia, Spain \\
  \email{alejandro.mus@uv.es}
  \and
  Observatori Astron\`omic, Universitat de Val\`encia, Parc Cient\'ific, C. Catedr\`atico Jos\'e Beltr\'an 2, 46980 Paterna, Val\`encia, Spain
  %\and 
  %Melmac-University, \email{gshumway@alf.de}
}

\date {Received  / Accepted}

\authorrunning{M\"uller+Mus}

\abstract
%context heading (optional)
{Imaging in radioastronomy is an ill-posed inverse problem. With the increasing sensitivity and telescopes capabilities, several strategies have been developed in order to solve this challenging problem. In particular, novel algorithms have been proposed recently using (constrained) nonlinear optimization and Bayesian inference.}
% aims heading (mandatory)
{The Event Horizon Telescope (EHT) Collaboration investigated the fidelity of their image reconstructions convincingly by large surveys solving the problem with different optimization parameters. This strategy faces a limitation for the existing methods when imaging the active galactic nuclei (AGN): large and expensive surveys solving the problem with different optimization parameters are time-consumptive. We present a novel nonconvex, multiobjective optimization modeling approach that gives a different type of claim and may provide a pathway to overcome this limitation.}
% methods heading (mandatory)
{To this end we used a multiobjective version of the genetic algorithm (GA): the Multiobjective Evolutionary Algorithm Based on Decomposition, or MOEA/D. GA strategies explore the objective function by evolutionary operations to find the different local minima, and to avoid getting trapped in saddle points.}
% results heading (mandatory)
{First, we have tested our algorithm (MOEA/D) using synthetic data based on the 2017 Event Horizon Telescope (EHT) array and a possible EHT + next-generation EHT (ngEHT) configuration. We successfully recover a fully evolved Pareto front of non-dominated solutions for these examples. The Pareto front divides into clusters of image morphologies representing the full set of locally optimal solutions. We discuss approaches to find the most natural guess among these solutions and demonstrate its performance on synthetic data. Finally, we apply MOEA/D to observations of the black hole shadow in Messier 87 (M87) with the EHT data in 2017.}
% conclusions heading (optional)
{MOEA/D is very flexible, faster than any other Bayesian method and explores more solutions than Regularized Maximum Likelihood methods (RML). We have done two papers to present this new algorithm: the first explains the basic idea behind multi-objective optimization and MOEA/D and it is used to recover static images, while in the second paper we extend the algorithm to allow dynamic and (static and dynamic) polarimetric reconstructions.}

\maketitle

\section{Introduction}
Very long baseline interferometry (VLBI) is a radio interferometric technique. The recorded signals at each antenna pair in the array are correlated gradually sampling the Fourier transform of the true image brightness distribution with Fourier frequencies determined by baselines projected on the sky plane. Imaging, i.e. the procedure of creating an image from sparsely sampled Fourier coefficients (visibilities), is a challenging ill-posed inverse problem.

Three main families of imaging algorithms in radioastronomy have been proposed: CLEAN-based~\citep{Hogbom1974, Clark1980, Bhatnagar2004, Cornwell2008, Rau2011, Mueller2022b}, maximum entropy-based/regularized maximum likelihood-based (RML)~\citep{Cornwell1985, Chael2016, Chael2018, Akiyama2017, Akiyama2017b, Mueller2022} and Bayesian-based~\citep{Arras2021, Broderick2020, Tiede2022}. Since the point spread function (PSF) has non-vanishing kernel (i.e. missing data), there is a not a unique image reconstruction. The solution to the unpenalized optimization problem is multivalued. This inherent degeneracy is addressed by the regularizer either by imposing support constraints manually (placing ``CLEAN windows'') in CLEAN or by adding a penalty term (RML) to the objective functional. However, in this way only one representative solution is recovered and does not reflect the problem of missing data, i.e. the fact that more than one model fits the visibilities and an unique mathematically ideal solution does not exist. Instead, for these methods the problem of missing data could be addressed by testing the reconstructions with different combinations of penalizations as done in \citet{eht2019d, eht2022c}. Alternatively, a third family, the Bayesian methods, impose a prior distribution and look for all possible images that fit the data, but the high computational cost makes them very slow and their require powerful machines. 

RML methods minimize the weighted sum of a data-fidelity term (ensuring proximity to the observed data) and a regularization term (ensuring simplicity and fidelity of the guess solution). Viable regularization terms that were used in frontline VLBI applications such as observations with the Event Horizon Telescope (EHT) include total flux and non-negativity constraints, smoothness assumptions (total variation, total squared variation), entropy functionals and sparsity priors. For more details, see the discussion in \citet{eht2019d}. More recently multiscalar penalization terms have been proposed \citep{Mueller2022, Mueller2022c}. Most RML techniques use local search techniques by quasi-Newton approaches. While the reconstructions are generally excellent, improving over CLEAN in resolution, accuracy and dynamic range in particular for very sparse data \citep[see e.g. the comparisons in][]{eht2019d, eht2022c, Mueller2022, ngehtchallenge}, this strategy comes with two major drawbacks. 

First, the landscape of the objective is highly complex. Hence, local optimization strategies could get easily trapped in local minima rather than reaching the global minimum. Since station-based gains have a priori known uncertainties (which can however be unbound in one direction in the case of uncharacterized telescope errors) that need to be self-calibrated during the imaging, current RML pipelines use the closure phases and closure amplitudes instead of the visibilities in a first round of self-calibration~\citep{eht2019d, eht2022c}. The hybrid imaging problem (i.e. the common reconstruction of the gains and the image in alternating imaging and self-calibration rounds) is non-convex. Similarly, the projection to the calibration independent closure quantities is non-convex as well and therefore, the use of gradient-descent based optimization strategies further is questioned, i.e. the optimization problem is multi-modal. Although, this issue may be addressed effectively by systematic tests of various starting models, regularization parameter combinations and reconstruction methods \citep[as done in][]{eht2019d, eht2022c}, a more global optimization strategy is desired.

Second, with a larger number of possible regularization terms that need to be combined to achieve an ideal image, it is unknown a-priori which selection of weighting parameters to choose. This problem is typically solved with a brute-force approach: a library of synthetic data is created that needs to be studied with every possible parameter combination (parameter survey). Only the parameter combination that passes several acceptance criteria will be used for the analysis of the real observational data. This procedure is tedious and time-consuming. Moreover, the process is poorly motivated. The set of test images could impact the topset-selection and thus the quality of the reconstruction. A multiobjective algorithm that evolves the subspace of non-dominated solutions in parallel and selects the ideal hyper-parameter array automatically is therefore needed.

Overall, applications to very sparse VLBI data sets such as in \citet{eht2019d, eht2022c} were successful in addressing both of the issues raised above related to the multi-modality and multi-objectivity of the problem by parameter surveys, combining the reconstructions by different methods and teams, and by extensive testing of the data set. This strategy allowed for strong indications of the fidelity of the ring-like appearance of the black hole shadow. In this work we build upon the success of such a survey strategy in identifying the fidelity of the recovered images, but look for a reformulation of the problem that allows for a faster, less time-consuming alternative. We present a novel imaging algorithm that provides an independent, alternative claim on the morphology of the image. Moreover, this algorithm may provide the possibility to accelerate parameter surveys. To this end, we present a novel formulation for RML-like problems to adapt to the multi-modality and multi-objectivity of the problem. All the solutions calculated in a parameter survey span a subspace of ``optimally-balanced'' (or more correctly, non-dominated solutions). Instead of computing this subspace by independent optimizations on a regular grid of coordinates and selecting the best representant (i.e. computing a parameter survey), we aim for the complete subspace as the result of our optimization procedure. This new modeling consists on solving a multiobjective optimization problem (MOP) where the objective function is a combination of the most used regularizing terms in the current RML methods \citep{Chael2016, Akiyama2017, Akiyama2017b, Chael2018}. We solve this problem by using the global search technique of genetic algorithms (GA), in particular we utilize the Multiobjective Evolutionary Algorithm Based on Decomposition (MOEA/D) algorithm \citep{Zhang2008}. In this way, we avoid to fall in one local minimum and we ensure more diversity on the solutions. We compute a set of candidate solutions that are non-dominated, i.e. optimal with respect to the corresponding regularization. Our strategy is similar in philosophy to parameter surveys for RML methods, but not equivalent since we use a different optimality concept. Instead of searching over several hyper-parameter combinations, we jointly evolve the solutions to all hyper-parameter combinations together, i.e. we speed up the exploration of hyper-parameters drastically by the genetic crosstalk of similar regularizer weight combinations. In a consecutive paper, we will extend the problem to capture dynamics (and then, to reconstruct ``movies'' instead of snapshots) and polarimetry.

On the one hand, in contrast to CLEAN and RML methods, we do not recover only one representative solution, but the hypersurface of all non-dominated (locally optimal) models. Every model in this surface of possible solutions corresponds to a specific regularizer weight combination that are all explored in parallel. This diversity of solutions represents the full set of possible image morphologies. To find the most natural solution within this hypersurface, we propose two different approaches: by the minimal distance to the ideal point and by looking for accumulation points. On the other hand, Bayesian methods are able to reconstruct a set of candidate solutions related to the multimodality of the problem as well, but the parameter space they explore makes their time performance to be low (they require quite a few hours or days of computation) for that scope, although there is considerable current progress in speeding up the performance \citep[e.g.][]{Knollmueller2019, Tiede2022} and Bayesian methods have consequently been applied even for large data sets \citep[e.g.][]{Arras2021, Tychoniec2022}. In contrast, the set of candidate solutions computed by MOEA/D does not have a Bayesian interpretation as a posterior distribution, and hence does not allow for an uncertainty quantification. \\
\\
The paper is divided as follows: In Sec.~\ref{sec:vlbibases} we describe the basics of VLBI measurements and imaging reconstruction using RML methods. In Sec.~\ref{sec:mop_sec} and Sec.~\ref{sec: moead} we give a short overview of multiobjective optimization and we introduce the terms that will be useful later, together with a global search technique used to solve this type of problems. We present the model of our problem in Sec.~\ref{sec: modelization}. Every solution of this problem is an optimal image. We test our algorithm and we discuss several points in Sec.~\ref{sec:synthetic_data} (synthetic data) and Sec.~\ref{sec:real_data} (real data). For all the tests, we have used self-calibrated data. In Sec.~\ref{sec:closure_only} we run the algorithm in not self-calibrated data and we study the importance of the initial point to constraint the problem when there is not a self-calibration model in the case of a sparse uv-coverage, while for a better uv-coverage, any extra constraint is needed to recover the intrinsic source structure. The main part of this first paper ends in Sec.~\ref{sec:summary}, which contains the conclusions of the work. Further appendices are included to avoid an unnecessary extension. 

\section{VLBI measurements and imaging}\label{sec:vlbibases}

An interferometer consists of a set of $T$ telescopes observing the same source at the same time. The signal recorded at two independent stations is correlated. This correlation product is called visibility $\mathcal{V}(u,v)$ with harmonic coordinates $(u,v)$ determined by the baseline of the antenna pair. The true sky brightness distribution $I(l,m)$ and $\mathcal{V}$ are related by the van-Cittert-Zernike theorem \citep{Thompson1994}:
\begin{align}
    \mathcal{V} (u, v) = \int \int I(l, m) e^{-2 \pi i (l u + m v)} dl dm \,.  \label{eq: vis}
\end{align}
With a full aperture, the true image could be recovered from the visibilities by an inverse Fourier transform. However, for VLBI only a sparse subsample of the Fourier coefficients is measured. We call the subsample of measured Fourier frequencies the uv-coverage. The measured visibilities along one pair of antennas (indexed with $i,j \in {1,2, ..., T}$) at a time $t$ is corrupted by station based gains $g_i$ and additional thermal noise $N_{i,j}$ specific to the baseline, such that the measured visibility on this baseline reads:
\begin{align}
    V(i,j,t) = g_i g_j^* \mathcal{V}(i,j,t) + N_{i,j}\,. 
\end{align}
Closure quantities are gain-independent quantities derived from the observed visibilities, i.e. the closure phases over a triangle of antennas $i,j,k \in \{1,2,...,T\}$ are:
\begin{align}
   \Psi_{ijk} = arg \left( V_{ij} V_{jk} V_{ki} \right)\,, 
\end{align}
and closure amplitudes over a rectangle of antennas $i,j,k,l \in \{1,2,...,T\}$ are:
\begin{align}
    A_{ijkl} = \frac{|V_{ij}| |V_{kl}|}{|V_{ik}| |V_{jl}|}\,.
\end{align}

For RML methods, we optimize not a single but several objective functionals \citep{Chael2016, Chael2018, Akiyama2017, Akiyama2017b, eht2019d, Mueller2022}. These include data fidelity terms that measure the fidelity of the guess solution to the observed data, e.g. the fit quality to the visibilities:
\begin{align}
    S_{\text{vis}} (\mathscr{V}, V) = \frac{1}{N_{\text{vis}}} \sum_{i=1}^{N_{\text{vis}}} \frac{|\mathscr{V}_i - V_i|^2}{\Sigma_i^2}\,,
\end{align}
with number of observed visibilities $N_{vis}$, visibilities of guess solution $\mathscr{V}_i$ and error $\Sigma_i$. Moreover, the fit to the amplitudes:
\begin{align}
    S_{\text{amp}} (\mathscr{V}, V) = \frac{1}{N_{\text{vis}}} \sum_{i=1}^{N_{\text{vis}}} \frac{(|\mathscr{V}_i| - |V_i|)^2}{\Sigma_i^2}\,,
\end{align}
to the closure phases:
\begin{align}
    S_{\text{cph}} (\mathscr{V}, V) = \frac{1}{N_{\text{cph}}} \sum_{i=1}^{N_{\text{cph}}} \frac{|\Psi_i(\mathscr{V}) - \Psi_i(V)|^2}{\Sigma_{\text{cph},i}^2}\,, \label{eq: cph}
\end{align}
and closure amplitudes:
\begin{align}
    S_{\text{cla}} (\mathscr{V}, V) = \frac{1}{N_{\text{cla}}} \sum_{i=1}^{N_{\text{cla}}} \frac{|\ln A_i(\mathscr{V}) - \ln A_i(V)|^2}{\Sigma_{\text{cla},i}^2}\,, \label{eq: lca}
\end{align}
could be used. Regularization terms measure the feasibility of the solution to fit the data with a model that is as simple as possible. Usual choices include a flux constraint $f$:
\begin{align}
    R_{\text{flux}}(I, f) = \norm{\int \int I(l,m)dldm-f}\,,
\end{align}
where $f$ is the total flux together with norm constraints:
\begin{align}
    &R_{l1}(I) = \norm{I}_{l^1}\,, \\
    &R_{l2}(I) = \norm{I}_{l^2}\,,
\end{align}
smoothness priors:
\begin{align}
    &R_{tv}(I) = \int \norm{\nabla I}\ dl\ dm \,,
    &R_{tsv}(I) = \sqrt{ \int \norm{\nabla I}^2\ dl\ dm }\,,
\end{align}
and an entropy functional:
\begin{equation}
    R_{\text{entr}}\left(I\right) = 
        \displaystyle{\int \int} I \ln\left(\dfrac{I}{M}\right) dl dm\,,
\end{equation}
where $M$ denotes the brightness distribution of a model image.\\

In RML methods these terms are added with corresponding weighting parameters $\alpha,\beta,...,\iota \in \mathbb{R}$ to create the full objective functional $\mathscr{F}$:
\begin{multline}
    \mathscr{F} = \alpha S_{\text{vis}} + \beta S_{\text{amp}} + \gamma S_{\text{clp}} + \delta S_{\text{cla}} \\
    + \epsilon R_{\text{flux}} + \zeta R_{l1} + \eta R_{l2} + \theta R_{tv} + \iota R_{tsv} + \kappa R_{\text{entr}}. \label{eq: rml}
\end{multline}
The joined minimization of $S_{\text{vis}}$, $S_{\text{amp}}$, $S_{\text{clp}}$, and $S_{\text{cla}}$ combines redundant information (e.g. the information encoded by the closure amplitudes is also encoded in the amplitudes), a potential weak point of forward modelling techniques. %While it has become common standard for RML applications to approach the problem in such a manner, especially within the EHT \citep{eht2019d, eht2022c}, \textbf{and the intensive testing demonstrated that this approach is strong enough to constrain the image morphologies effectively,} the use of figure-of-merits may be poorly motivated due to the correlation between redundant data terms. 
In Sec.\ref{sec:closure_only} we show how our algorithm works with terms only closure dependent.

\section{Multiobjective Optimization (MOP)}
\label{sec:mop_sec}

A multiobjective (minimization) optimization problem (or MOP) in its general form can be stated as \citep{Pardalos2017}
\begin{problem}[MOP standard form]
  \begin{equation*}
  \label{prob:mop}
  \tag{$\text{MOP}$}
    \begin{aligned}
      & \underset{x\in D}{\text{min}}
      & & F\left(x\right) := \left(f_1\left(x\right),\ldots,f_n\left(x\right)\right),\\
      & \text{subject to}
      & & x\in D\subset\RR^m,\\
    \end{aligned}
  \end{equation*}
\end{problem}
where $D$ is the decision space, $\RR^n$ is the space of objectives, $F:D\longrightarrow\RR^n$ is the vector-valued multiobjective optimization functional whose each component $f_i:D\longrightarrow\RR,\ i=1\ldots,m$ is an objective functional. The feasible set $D\subset\RR^n$ is generally expressed by a number of inequality constraints $D=\{x\in\RR^n\ | \ c_i\left(x\right)\leq 0, i=1,\ldots,n\}$. This is the setting of our work.

It is common to find discrepancies between objectives, meaning that there does not exist a single point $x\in D$ that minimizes (maximizes) all the $f_i$ simultaneously \citep{Pardalos2017}. Therefore, the goal of the MOP is to find the best compromise among all of them. This compromise is defined by means of a special set of solutions: the Pareto front $\pfront$ \citep{Pardalos2017}.

The Pareto front consists of the Pareto optimal (non-dominated) solutions in the space of objectives. A point $x^*\in D$ is called Pareto optimal (non-dominated) if there is no point $x\in D$ such that $f_i\left(x\right)\leq f_i\left(x^*\right),\ \forall i=1,\ldots,m$ and $f_j\left(x\right)<f_j\left(x^*\right)$ for at least one $j=1,\ldots,m$. In other words, we call a point $x^* \in D$ Pareto optimal if the further optimization in one objective automatically has to worsen another objective functional.

In general, $\pfront$ cannot be found analytically. In particular, most of the efforts are devoted to approximate $\pfront$ or to identify characteristic members \citep{Pardalos2017}. The selection of these represent members should be carefully done to avoid unfeasible long running times or small diversity among the solutions. We also like to point out that the Pareto front represents a novel approach in VLBI to the multiple data and regularization terms discussed in Sec. \ref{sec:vlbibases}. With the weighted sum approach (Eq. \eqref{eq: rml}) with varying parameter combinations $\{ \alpha, \beta, ..., \iota\}$ we calculate a hypersurface as well that effectively approximates the Pareto front.

The Pareto front is bounded by the nadir and the ideal objective vectors~\citealp[see for instance][]{Pardalos2017}. While the nadir is not used in this work, and therefore, we refer to bibliography for a clear definition, the ideal objective vector is the element $l=\left(l_1,\ldots,l_m\right)$ on $D$ such that each component $l_i$ is computed by (compare also Fig. \ref{fig: pareto_scheme}): 
\begin{equation}\label{eq:ideal_point}
    l_i = \displaystyle{\inf_{x\in\pfront}}f_i\left(x\right),\ i=1,\ldots,m\,.
\end{equation}
In this work, we use this vector to define a metric used to return one representative image, despite all the images belonging to the Pareto Front are also equally valid solutions.

Among all the possible strategies used to approximate the Pareto front \citep[we refer to the recent summary][for a comprehensive overview]{Sharma2022}, the so-called multiobjective evolutionary algorithms (MOEAs) have been found to be efficient approaches. In this work, we have used MOEA Based on Decomposition (MOEA/D)~\citep{Zhang2008, Li2009}. This technique first obtains $\pfront$ by solving a set of scalar functionals associated to the objectives in a collaborating manner via an evolutionary algorithm. This cooperative strategy allows to handle large scale optimization problems by decomposing into smaller scale subproblems~\citep{Tsurkov2001}. MOEA/D has high search ability for continuous, combinatorial and multiobjective optimization. MOEA/D has a lower complexity than other algorithms. It is out of the scope of this paper to do a comparison of algorithms for solving MOP. We refere, for instance, to~\cite{Xin2019}. %such as NSGA-II, but it performs similarly in terms of solution quality \citep{Zhang2008}.

\section{MOEA/D} \label{sec: moead}
Nonconvex problems, as Prob.~\eqref{prob:mop_ours}, generally have more than one optimal solution. Such solutions are called \textit{local optimal solution}. Gradient or Hessian based algorithms are questionable in such type of problems, because they are only able to return the first local solution they find. We refer to \citet{Mus2023} for a longer discussion on the initial point dependence in noncovex problems. In this Section we summarize a global search strategy called MOEA/D first proposed in \citet{Zhang2008} that overcomes this issue. For more details we refer to \citet{Zhang2008, Li2009, Xin2019}.\\

MOEA/D solves the optimization problem with a genetic algorithm. Genetic algorithms are inspired by natural evolution. At every generation a population of solutions is created from the preceding generation. We calculate the evolution from one generation to the next generation by genetic operations, i.e. by random mutation of the single representants (genes) in the population and mating (i.e. mixing) of randomly selected parents in the parent-generation. 

In MOEA/D the problem is decomposed in single problems either by a Tchebycheff decomposition or by a weighted sum decomposition. For this work we focus on the weighted sum decomposition due to its philosophical similarity to established RML reconstructions in VLBI \citep{Johnson2017, Akiyama2017, Chael2016, Chael2018}. We define weight arrays $\lambda^1 = \{\lambda_0^1, \lambda_1^1, ..., \lambda_m^1\}, \lambda^2=\{\lambda_0^2, \lambda_1^2, ..., \lambda_m^2\}, ...$ that are related to the objective functionals $f_1, f_2, ..., f_m$. Every weights array is normalized, i.e.:
\begin{align}
&\lambda_i^j \in [0,1],\ \forall i,j\,, \\
&\sum_{i=1}^m \lambda_i^j = 1,\ \forall j\,. \label{eq: sum_weights}
\end{align}
Prob.~\eqref{prob:mop} is decomposed into solving single optimization problems by a weighted sum approach:
\begin{align}
x^j \in \mathrm{argmin}_{x} \sum_{i=1}^m \lambda_i^j f_i(x)\,.
\end{align}
The non-dominated single solutions in $\{x^j\} \in D\subset\RR^n$ approximate the Pareto front. The optimization is done with a genetic algorithm that interchanges information in between several genes in one population at every iteration. For the details we refer to \citet{Zhang2008} and \citet{Li2009}. In a nutshell, we define closest neighborhoods $U_B(\lambda^j)$ around every weights array $\lambda_j$. The update step consists roughly of the following substeps done for every $j$:
\begin{enumerate}
    \item We select two random indices $k,l$ out of the neighborhood $U_B(\lambda^j)$.
    \item We generate a new solution $y^j$ by genetic operations from $x^k$ and $x^l$, i.e. by random mutation and crossover among different candidates.
    \item We update all the solutions in the neighborhood, i.e. for all indices $n \in U_B(\lambda^j)$ we set $x^n=y$ if $\sum_{i=1}^m \lambda_i^n f_i(y^j) < \sum_{i=1}^m \lambda_i^n f_i(x^n)$.
    \item We find the non-dominated solutions.
\end{enumerate}
MOEA/D therefore evolves the population at every point in the Pareto front at the same time. Moreover, it preserves diversity since isolated neighborhoods are protected from each other.

\section{Modelization of the problem} \label{sec: modelization}

To model the problem, we have chosen seven objective functionals:
\begin{align}
    & f_1:=\alpha S_{vis}+\beta S_{amp}+\gamma S_{clp}+\delta S_{cla}+\zeta R_{l1},\\
    & f_2:=\alpha S_{vis}+\beta S_{amp}+\gamma S_{clp}+\delta S_{cla}+\theta R_{tv},\\
    & f_3:=\alpha S_{vis}+\beta S_{amp}+\gamma S_{clp}+\delta S_{cla}+\tau R_{tsv},\\
    & f_4:=\alpha S_{vis}+\beta S_{amp}+\gamma S_{clp}+\delta S_{cla}+ \eta R_{l2},\\
    & f_5:=\alpha S_{vis}+\beta S_{amp}+\gamma S_{clp}+\delta S_{cla}+\epsilon R_{flux},\\
    & f_6:=\alpha S_{vis}+\beta S_{amp}+\gamma S_{clp}+\delta S_{cla}+\kappa R_{entr},\\
    & f_7:=\alpha S_{vis}+\beta S_{amp}+\gamma S_{clp}+\delta S_{cla}.
\end{align}

Therefore, the nonconvex multiobjective problem to be solved is 
\begin{problem}[MOP for imaging reconstruction]
  \begin{equation*}
  \label{prob:mop_ours}
  \tag{$\text{MOP-MOEA/D}$}
    \begin{aligned}
      & \underset{x\in D}{\text{min}}
      & & F\left(x\right) := \left(f_1\left(x\right),\ldots,f_7\left(x\right)\right),\\
      & \text{subject to}
      & & x\in \RR^{\text{Npix}}_{+},\\
    \end{aligned}
  \end{equation*}
\end{problem}

This modelization is flexible and it is easy to include new regularization functionals. Due to Eq. \eqref{eq: sum_weights}, the weighted sum decomposition with weights array $\{\zeta, \epsilon, \eta, \theta, \tau, \kappa, 1-\zeta-\epsilon-\eta-\theta-\tau-\kappa \}$ is equivalent to Eq. \eqref{eq: rml}. Hence, every item in the Pareto front corresponds to the optimal solution to a single hyper-parameter combination (thus replacing parameter surveys). Moreover, this selection of the objective functionals ensures that the optimization is compatible with the data in every optimization direction (i.e. every $f_i$ assures fit quality to the data). In contrast to parameter surveys that were proposed for RML methods, but are numerically unfeasible and poorly motivated, all solutions corresponding to a specific regularizer weight combination $\lambda^j$ are evolved together. Due to the genetic mixing of neighboring solutions, MOEA/D shares information between the solutions for similar weight arrays $\lambda^j$ and improves in this regard over parameter surveys that calculate the solutions independently. However, we have to mention that while the similarity to parameter surveys constitutes the main motivation behind the application of MOEA/D for VLBI imaging, Pareto-optimality is a slightly different optimization concept.

Although we are using the combined objectives $f_1, ..., f_7$ for the MOEA/D, during postprocessing we examine the front related to the penalty terms $R_{l1} \propto f_1-f_7, R_{tv} \propto f_2-f_7, ...$. % We show in Fig.~\ref{fig: crescent_ngeht} the Pareto front for a synthetic ground truth crescent model observed with the EHT + ngEHT configuration. 
For more details we refer to Sec. \ref{sec: ngeht}, were we show examples of different Pareto fronts. We plot the Pareto front in the first row as projections onto the three-dimensional domain. In the second row we present the same front, but with different projections. Every single point in these plots corresponds to an image recovered for a specific weighting combination. When inspecting the front we can identify several disjoint clusters. These clusters demonstrate several image morphologies that become visible when changing the weight parameter combinations. The image diversity within one cluster on the contrary is small. We find the clusters by a standard clustering algorithm. First we renormalize every axis such that the values are in the range $[0,1]$. Then, we find for every point the respective neighbors (where a point is classified as a neighbor when the relative distance between them in the seven-dimensional space spanned by $[f_1, ..., f_7]$ is larger than some threshold). Finally, we classify all data points that are connected by a path of neighbors as being part of the same cluster. 

We like to mention that every solution in the Pareto front is an optimal solution with respect to the multiobjective optimization problem. Mathematically there is no preferred solution. All image morphologies in the different clusters are mathematically reasonable solutions that fit the data. Therefore, the Pareto front is the main output of MOEA/D representing the fact that there is not a single preferred image due to missing Fourier coefficients, and presents an illustration of possible image features, similar in philosophy to draws from the posterior distribution for Bayesian algorithms or the survey of images produced in a parameter survey. However, it is common standard in VLBI imaging to select one image that is most natural with respect to scientific perception. We present two strategies of finding the most natural image among all optimal solutions. 

\begin{figure}
    \centering
    \includegraphics[width=0.45\textwidth]{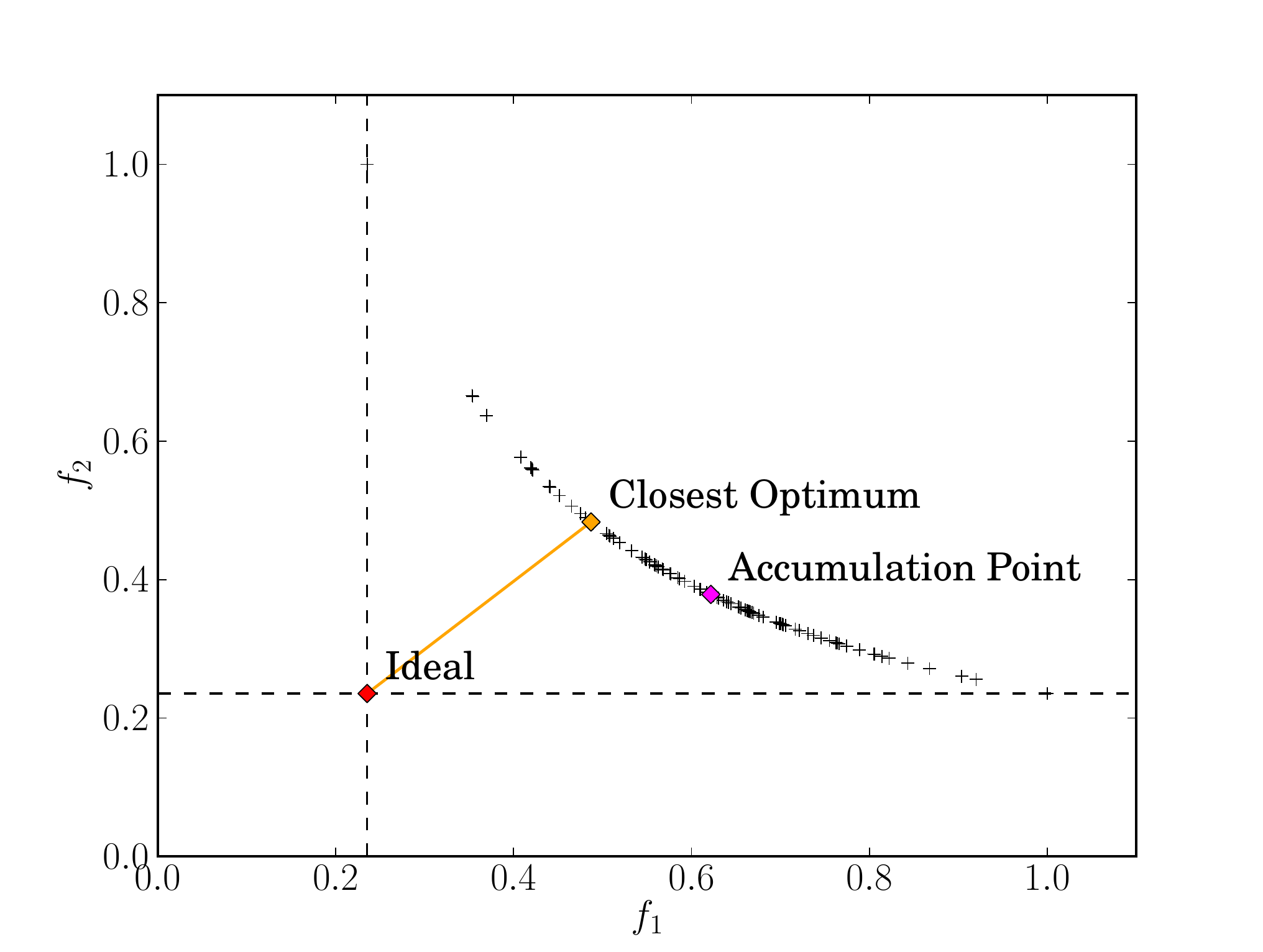}
    \caption{Schematic illustration of the Pareto front for a two given functions $f_1$ and $f_2$, its ideal point and two concepts for the most natural preferred points.}
    \label{fig: pareto_scheme}
\end{figure}

These two strategies are illustrated in Fig. \ref{fig: pareto_scheme}. First we can define the ideal point following Eq.~\eqref{eq:ideal_point}. As the ideal point is the point that would be optimal in every objective, it can be found by the cross-section of the minimum in every single objective (see Fig. \ref{fig: pareto_scheme}). The ideal, however, has no physical meaning, since it is not a solution of the Prob.~\eqref{prob:mop_ours}. A natural choice for a single image instead would be the recovered solution in the Pareto front that has the shortest (euclidean) distance to the ideal point. 

As a second strategy, we can look for accumulation points. Every cluster of solutions represents a locally optimal image morphology. We assume that most solutions may cluster among the most natural solutions, while the edges of the parameter space with more exotic solutions are less sampled. To find a natural representative image, we therefore look for the points with the highest number of close neighbors, the accumulation point of the cluster. 

The traditional MO theory lacks a standard methodology to address the challenge of solution selection, which entails the identification of a representative solution from among the entire set of solutions comprising the Pareto front. Although a few strategies have been put forth in the literature,~\citep[for instance][]{Schwind2014,Schwind2016}, there is no consensus on an optimal approach. In this paper, we introduce two distinct criteria for solution selection: the accumulation point and the closest image to the ideal. Notably, any method that yields an image belonging to the Pareto front is a valid method of selecting a mathematical solution for the given problem.

\section{Verification on synthetic data}\label{sec:synthetic_data}

\subsection{Synthetic Data}
We have tested our algorithm on the four synthetic geometric models (double, crescent, disk and ring) that were used for the imaging verification by the EHT in 2017 \citep{eht2019d}. For each model, two observational data sets were generated, one based on the EHT array of 2017 (coverage and thermal noise-level of the 2017 observations of M87 on 05. April) and the other based on a possible EHT + ngEHT configuration \citep{ngehtchallenge} including ten additional stations, a quadrupled bandwidth and an enhanced frequency coverage. We simulate synthetic data at $230\,\mathrm{GHz}$ and add thermal noise according to the expected systematic noise levels used for \citet{ngehtchallenge}. %The uv-coverages for these two arrays are plotted in Fig. \ref{fig:synth_uvcov}. 

The geometric models are the same that were used for imaging verification in \citet{Mueller2022} and \citet{eht2019d}. The size of the ring, crescent, and disk mimics the size of the famous black hole shadow in M87 \citep{eht2019d}. The ring and crescent have a radius of $22\,\mu\mathrm{as}$, the disk a diameter of of $70\,\mu\mathrm{as}$. The double image mimics a completely different image structure comparable to the features of 3C279 as seen by the EHT \citep{Kim2020}. All geometric models are normalized to a total flux of $0.6\,\mathrm{Jy}$. Moreover, we blur the geometric models with a $10\,\mu\mathrm{as}$-blurring kernel before synthetically observing them to avoid sharp edges.

% \begin{figure*}
%     \centering
%     \includegraphics[scale=0.3]{uv-coverage/uvcov_hi101.pdf}
%     \includegraphics[scale=0.3]{uv-coverage/ngeht_uvcov.pdf}
%     \caption{uv-covarage for M87 April 11 2017 EHT (left panel) and EHT + ngEHT (right panel) array.}
%     \label{fig:synth_uvcov}
% \end{figure*}

\begin{figure}
    \centering
    \includegraphics[scale=0.3]{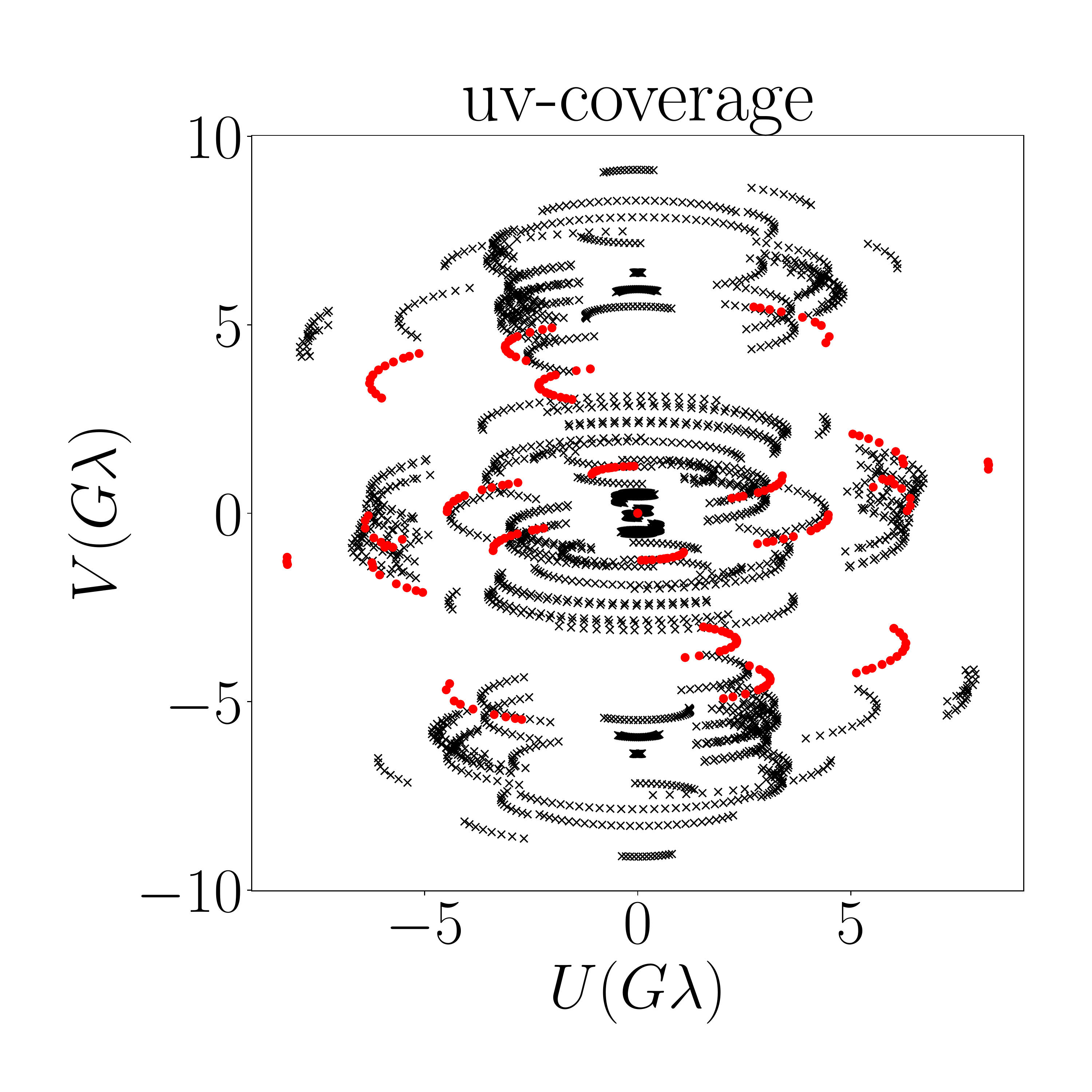}
    \caption{uv-coverage for M87 April 11 2017 EHT (red dots) and ngEHT at 345\,GHz (black crosses).}
    \label{fig:uv_cov}
\end{figure}

\subsection{EHT + ngEHT array} \label{sec: ngeht}
Figure~\ref{fig:uv_cov} depicts the uv-coverage of the EHT + ngEHT array. Red points correspond to the EHT baselines and black crosses to the ngEHT array. Crosses and and points together form the full EHT + ngEHT array. The uv-plane is less sparse, thus leading to improved constraints on the inverse problem, which remains ill-posed. To mimic the uncertainty in the phases we do the reconstruction from the amplitudes and closure quantities only as was proposed in \citet{Chael2018, Mueller2022} and applied in \citet{eht2019d, eht2022c}. In particular we use the data term combination $\alpha=0$ and $\beta=\gamma=\delta=1$. The weights were chosen on a grid with ten steps in every dimension, giving rise to $3003$ parameter combinations that suffice Eq. \eqref{eq: sum_weights}. We use a genetic algorithm with 4000 iterations. Moreover, we set the hyperparameters related to the genetic operations to the default values proposed in \citet{Li2009}. For more details on the optimal choice of the genetic parameters (e.g. mating probability, random mutation size and number of iterations) we refer to Appendix \ref{sec: parameter_survey}. Moreover, we add an overall scaling to every penalty term that was found to be ideal. For more details we refer to Appendix \ref{sec: parameter_survey}. We select a $40\,\mu\mathrm{as}$ Gaussian as a prior. The entropy functional is computed relative to this prior image. Moreover, we have to choose an initial population for MOEA/D. Instead of starting from a pre-imaging model, we start the MOEA/D algorithm from a random distribution. Every gene in the initial population is drawn independently from a uniform distribution with values between $0\,\mathrm{Jy}$ and $1\,\mathrm{Jy}$. Finally, all the initial genes in the initial population are rescaled to the respective compact flux.

\begin{figure}
    \centering
    \includegraphics[width=0.5\textwidth]{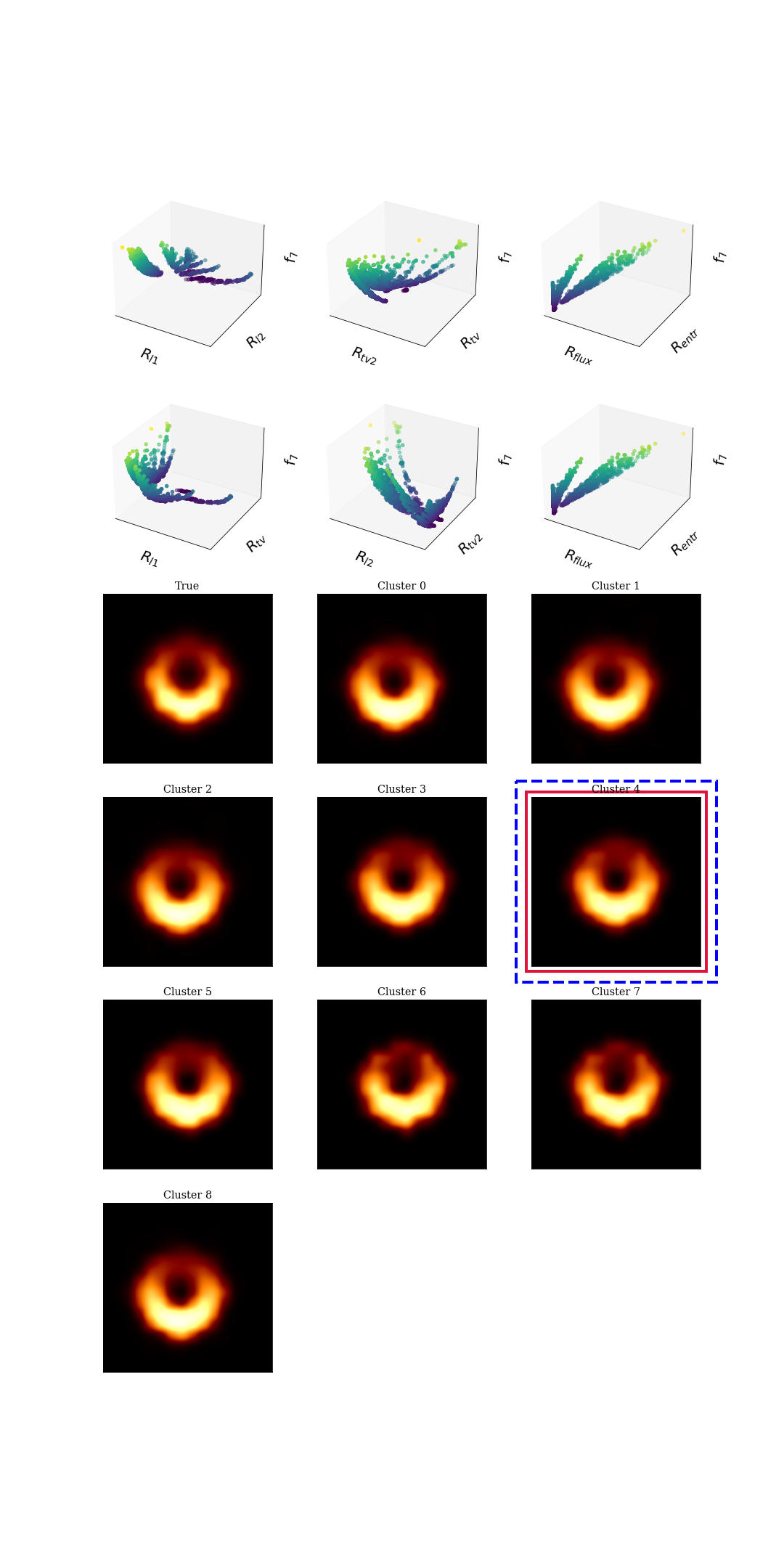}
    \vspace{-2cm}
    \hspace{-2cm}
    \caption{Pareto front (first two top rows) and solution clusters (following rows) for the crescent case using the EHT + ngEHT array. The top left panel of the solution clusters shows the true image. The Pareto front is a seven-dimensional hyper-surface. We illustrate the Pareto front by six projections. The six projections show the correlation between two regularizers and their values with respect to $f_7$ (data fidelity term only functional). The bluer the points are, the lower value for $f_7$. The family of solution can be grouped in 8 clusters. Red box surrounding the cluster indicates the preferred solution by the accumulation point strategy, while blue box is the one closest to the ideal.}
    \label{fig: crescent_ngeht}
\end{figure}

\begin{figure}
    \centering
    \hspace{-2.15cm}
    \includegraphics[width=0.55\textwidth]{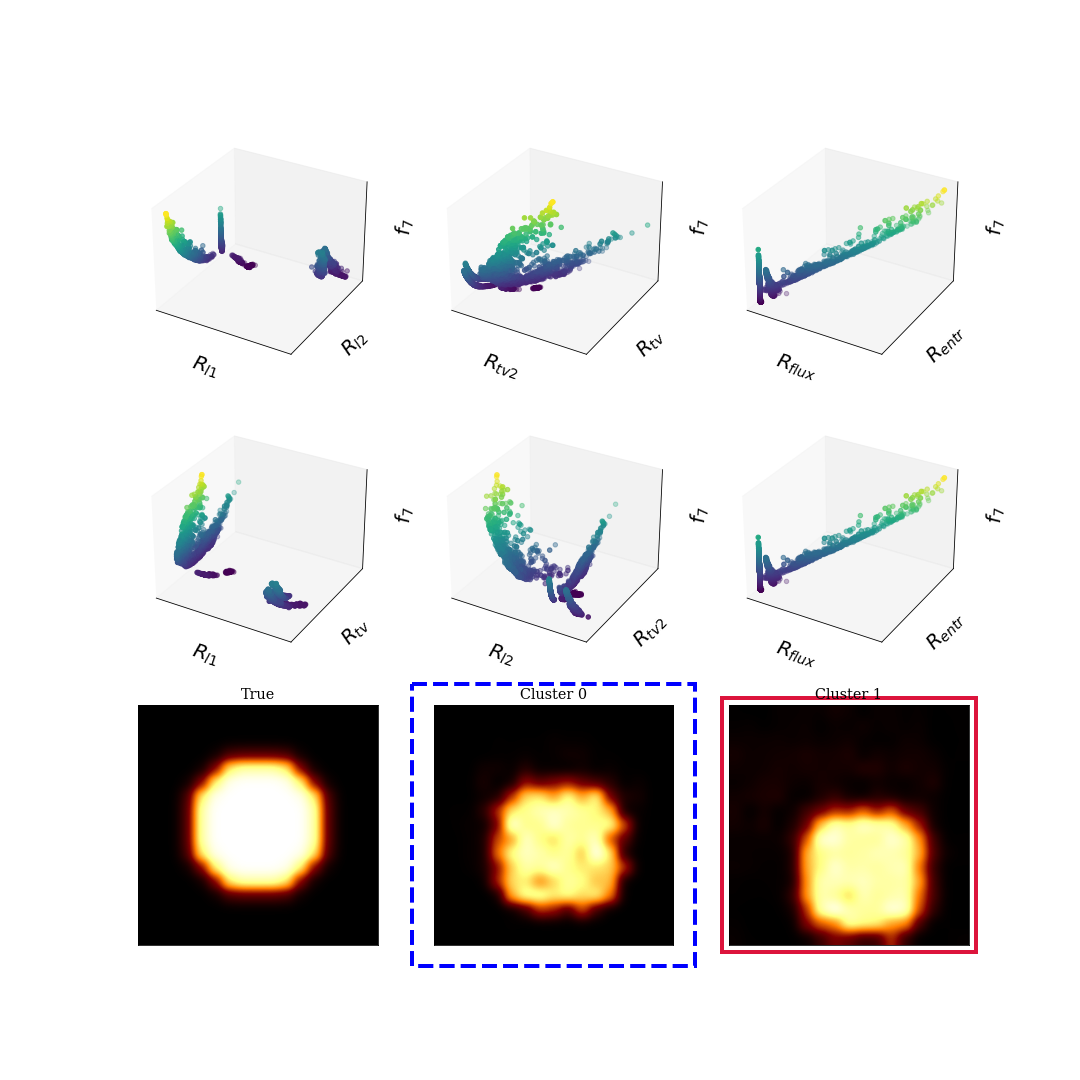}
    \vspace{-1cm}
    \hspace{-2.15cm}
    \caption{Same as Fig.~\ref{fig: crescent_ngeht} but for the disk model.}
    \label{fig: disk_ngeht}
\end{figure}

\begin{figure}
    \centering
    \hspace{-2.15cm}
    \includegraphics[width=0.55\textwidth]{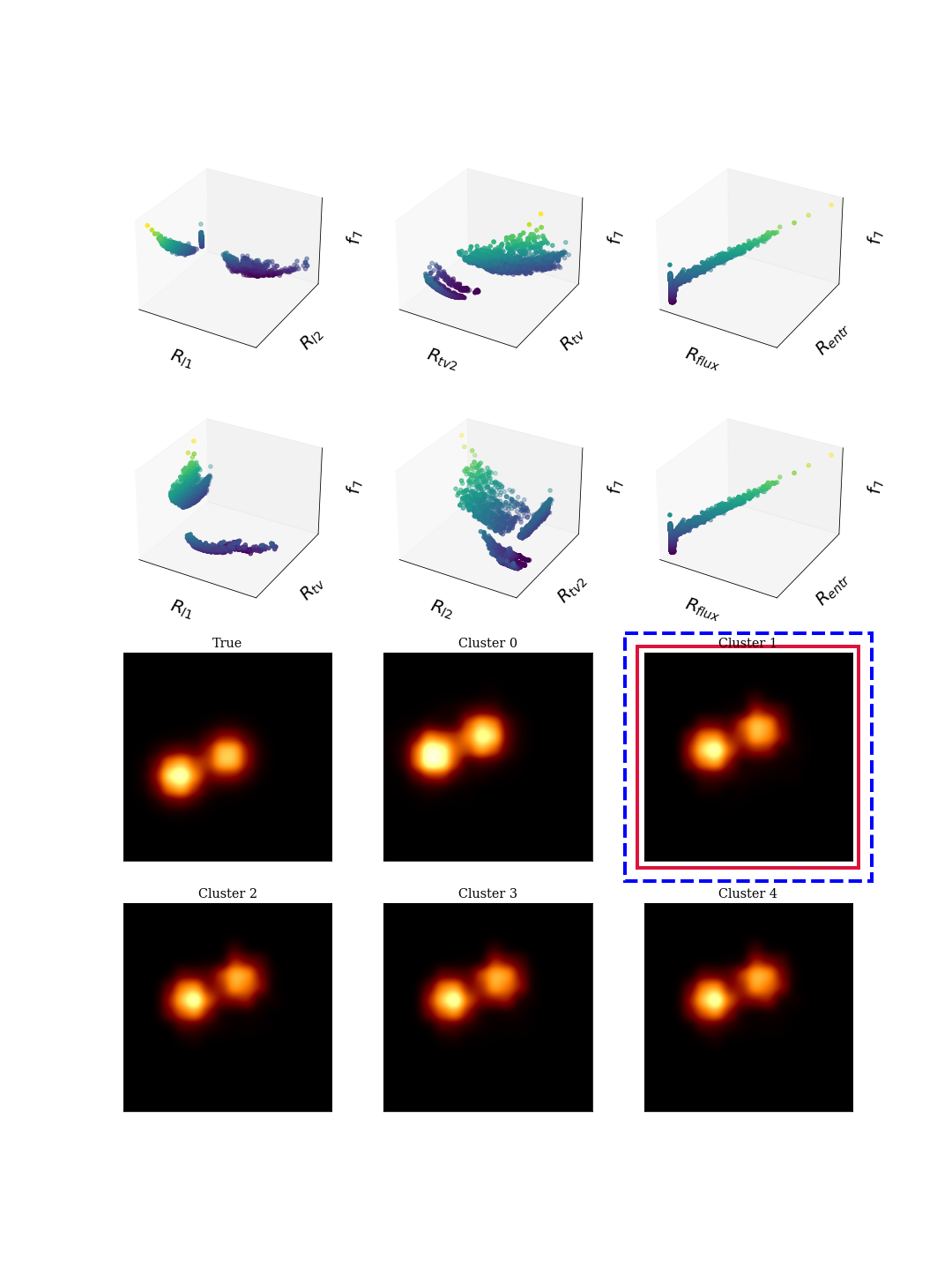}
    \vspace{-2cm}
    \hspace{-2.15cm}
    \caption{Same as Fig.~\ref{fig: crescent_ngeht} but for the double model.}
    \label{fig: double_ngeht}
\end{figure}

\begin{figure}
    \centering
    \hspace{-2.15cm}
    \includegraphics[width=0.55\textwidth]{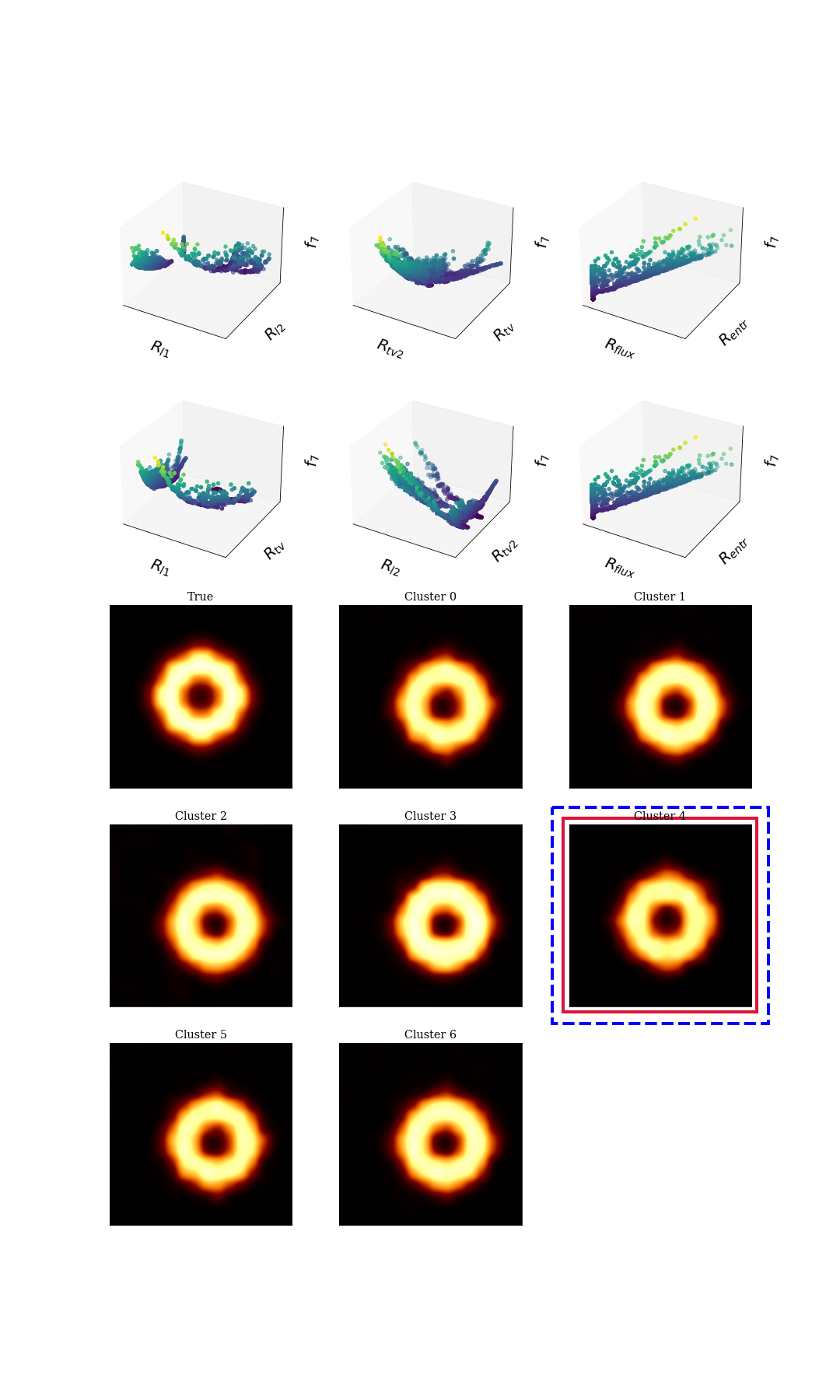}
    \vspace{-2cm}
    \hspace{-2.15cm}
    \caption{Same as Fig.~\ref{fig: crescent_ngeht} but for the ring model.}
    \label{fig: ring_ngeht}
\end{figure}

We show our results for the four geometric models in Fig. \ref{fig: crescent_ngeht}, Fig. \ref{fig: disk_ngeht}, Fig. \ref{fig: double_ngeht} and Fig. \ref{fig: ring_ngeht}. The Pareto front is a hypersurface in a seven-dimensional space (six penalty terms and one combined data term). We plot the front as a series of three projection into the three dimensional space (respectively two penalty terms and the data term) in the top row panels. As a first observation we see that the entropy, total flux constraint and data term are strongly correlated since a wrong total flux or source size worsens the fit to the observed amplitudes. 

When inspecting specific fronts, we see that $R_{l1}$ and $R_{l2}$, as well as $R_{tv}$ and $R_{tsv}$ seem to be correlated. That is reasonable as these regularization terms promote a similar prior information (sparsity and smoothness) respectively. In the second row panels we present the same Pareto front, but this time combine the terms with conflicting assumptions (smoothness versus sparsity), i.e. the various prior assumption that we aim to balance with an RML method. As expected we observe an anti-correlation in most cases. The Pareto front represents all optimal balances along this line of conflicting assumptions. The front divides into a varying number of clusters. The diversity of the images within every cluster is small, but the diversity from one cluster to the next cluster is significant. In the lower panels we show the ground truth image (top left) and a single representant (the accumulation point) of every cluster. We show with a red box the solution that is preferred by the accumulation point selection criterion, i.e. the one that has the largest number of close neighbors. Moreover, we show with a blue box the solution that is preferred by the closest optimum criterion. The two criteria coincide except for the disk, but in all cases select a reasonable reconstruction. For all four geometric models, the reconstruction is quite successful. The image features are recovered very well, although MOEA/D seems to prefer slightly blurred reconstructions. 

\subsection{EHT array}

Red points in Fig.~\ref{fig:uv_cov}, shows the uv-coverage of the supermassive black hole Messier87 (M87) during the 2017 EHT observation campaign. This poor coverage makes the imaging more challenging, in particular since we are limited to the closure quantities. This results in an harder optimization problem as demonstrated by the enhanced diversity in the population in the case of GA. The reconstruction was done with 5000 iterations and a random starting point. Moreover, we used the same genetic parameters as we used for the EHT + ngEHT data sets.

\begin{figure}
    \centering
    %\hspace{-2.15cm}
    \includegraphics[width=0.5\textwidth]{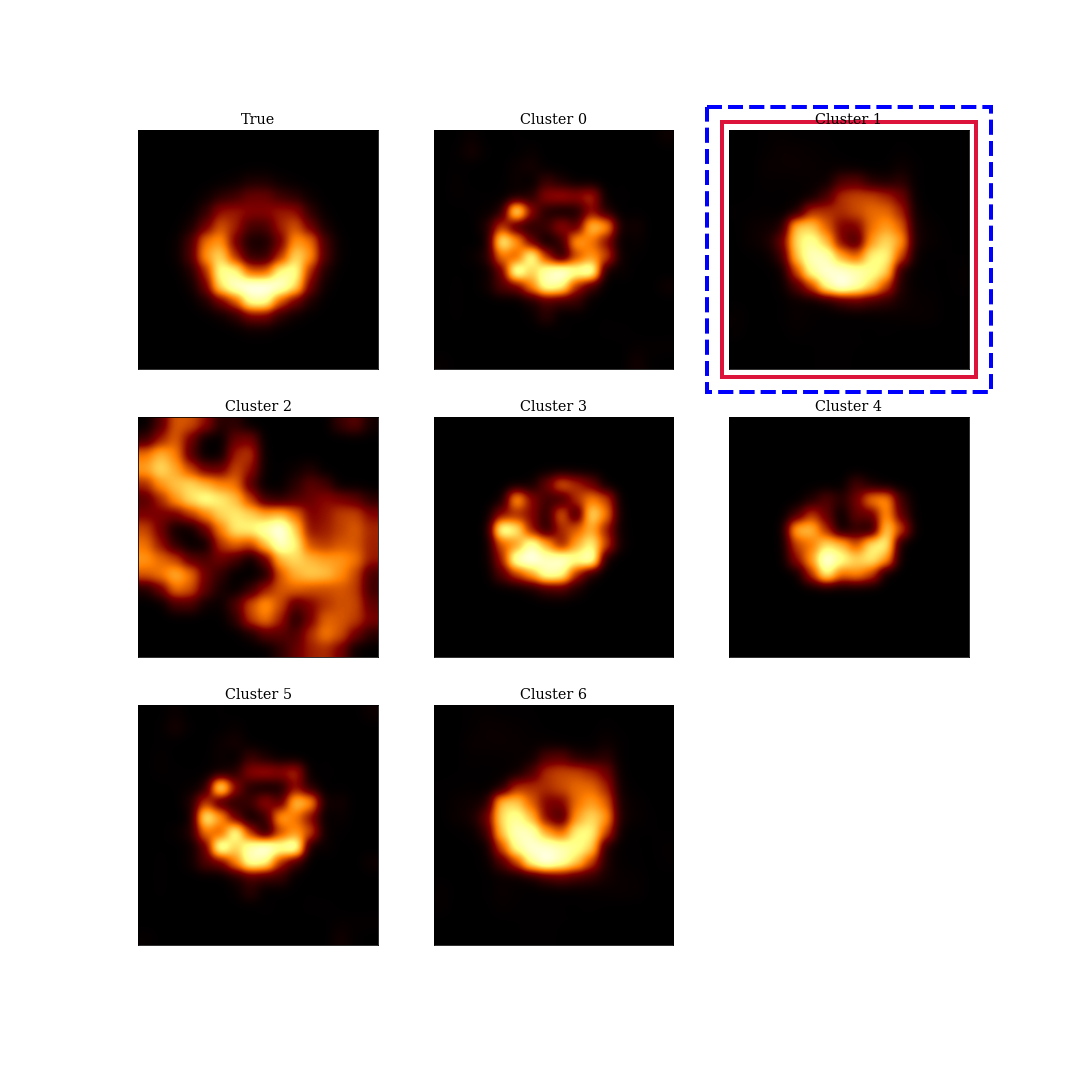}
    \vspace{-1cm}
    \caption{Solution clusters for the crescent case using the EHT array. The top left panel shows the true image. The red box surrounds the cluster indicated by the accumulation point strategy, while the blue box is the one closest to the ideal.}
    \label{fig: crescent_eht}
\end{figure}

\begin{figure}
    \centering
    %\hspace{-2.15cm}
    \includegraphics[width=0.5\textwidth]{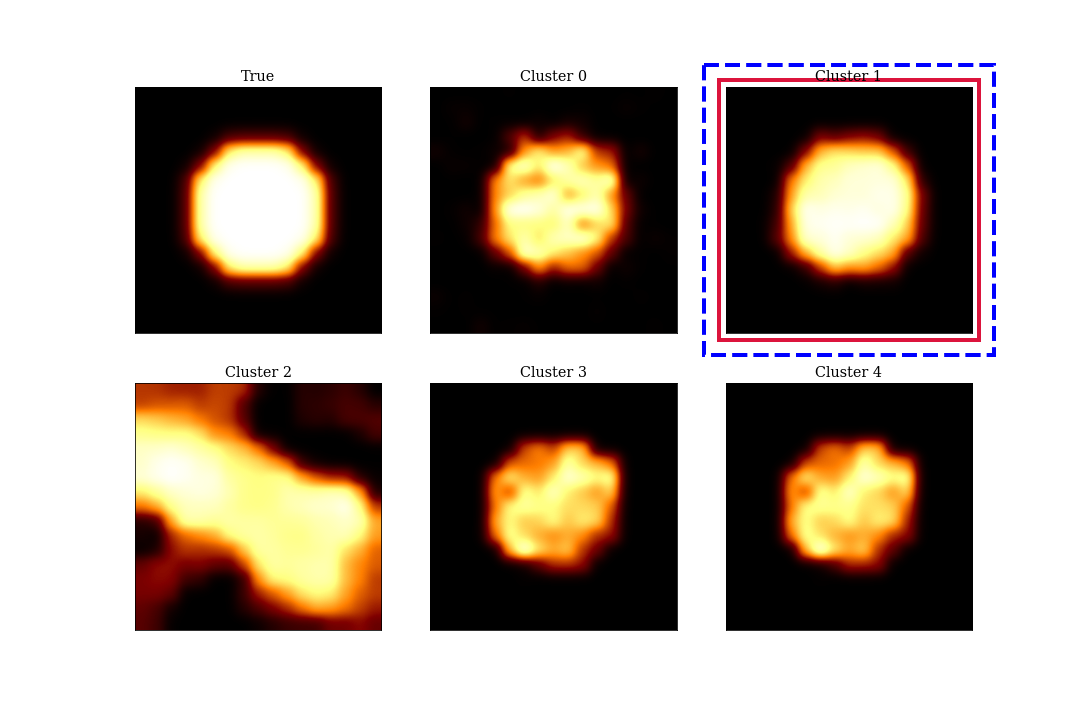}
    \vspace{-1cm}
    %\hspace{-2.15cm}
    \caption{Same as Fig.~\ref{fig: crescent_eht} but for the disk model.}
    \label{fig: disk_eht}
\end{figure}

\begin{figure}
    \centering
    \hspace{-2.15cm}
    \includegraphics[width=0.55\textwidth]{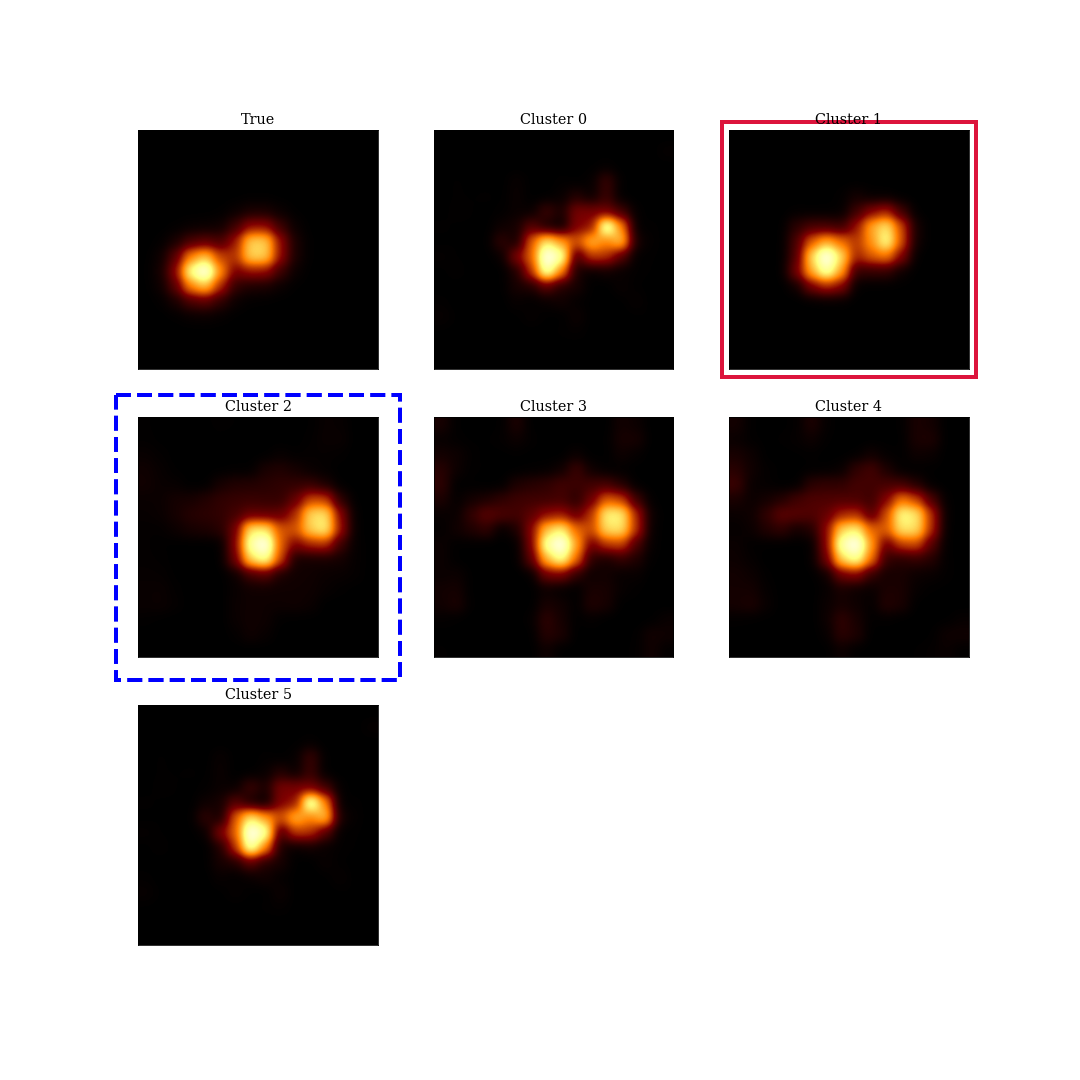}
    \vspace{-1cm}
    \hspace{-2.15cm}
    \caption{Same as Fig.~\ref{fig: crescent_eht} but for the double-source model.}
    \label{fig: double_eht}
\end{figure}

\begin{figure}
    \centering
    \hspace{-2.15cm}
    \includegraphics[width=0.55\textwidth]{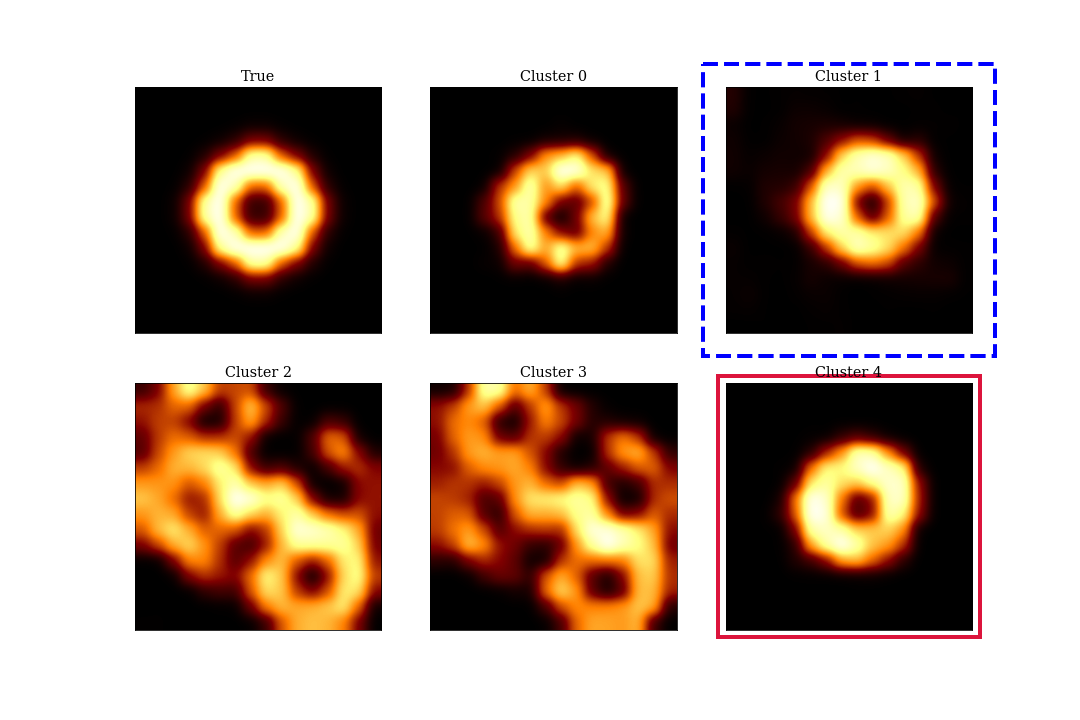}
    \vspace{-1cm}
    \hspace{-2.15cm}
    \caption{Same as Fig.~\ref{fig: crescent_eht} but for the ring model.}
    \label{fig: ring_eht}
\end{figure}

We show the recovered clustered images for all for geometric models in Fig. \ref{fig: crescent_eht}, Fig. \ref{fig: disk_eht}, Fig. \ref{fig: double_eht} and Fig. \ref{fig: ring_eht}. The recovered images show a wide variety of image morphologies. There are reasonable well (but slightly blurred) reconstructions (e.g. cluster 1 and 6 for the crescent or cluster 1 and 4 for the ring). Moreover, we observe over-resolved reconstructions (e.g. cluster 0 and cluster 4 for the crescent or cluster 0 and cluster 5 for the double structure). Since every point in the Pareto front has a one-to-one correspondence to a specific weight vector combination $\{ \lambda_i^j \}$, we can investigate which solutions are causing the over-resolved structural patterns. As expected these solutions are at the edge of the Pareto front with a dominating sparsity term. Finally, we find some clusters that show phantom repetition of the same structures (e.g. cluster 2 and 3 for the ring or cluster 2 for the disk). These secondary phantom images are not unusual for image reconstructions due to a combination of visibility phase uncertainty and
poor uv-coverage. This issue is addressed in RML algorithms either by surveying the correct balancing of various regularization terms \citep{eht2019d, eht2022c} or by a multiscale based hard thresholding \citep{Mueller2022}. An analysis of the reconstructions in MOEA/D shows that these solutions are related to the unpenalized reconstruction. That can be well explained by the dirty beam. In particular, the cluster 2 reconstruction of the disk example resembles the dirty image, i.e. the unpenalized reconstruction converges to the dirty image as the easiest solution that fits the data.

For all four geometric models, we select a well cluster of reconstructed images by the accumulation point strategy and the closest neighbor strategy. Hence, these strategies may give rise to a completely data-driven image reconstruction without the need of parameter surveys. Among all optimal images, we select the best one by looking for the image with the most close neighbors in the Pareto front.

\section{Real data}\label{sec:real_data}

We have applied our algorithm to real data taken during the 2017 EHT campaign (Papers I-VI). We have reconstructed images of M87 using the uvfits files available in the official web page of the EHT\footnote{The full link: \url{https://eventhorizontelescope.org/for-astronomers/data}}.

To obtain the image, we have used the best parameter setting discussed in Sec.~\ref{sec:synthetic_data}. For this work, we have considered three scenarios:
\begin{itemize}
    \item Scenario A: random starting point and random prior,
    \item Scenario B: random starting point and a Gaussian prior
    \item Scenario C: the image with less $\chi^2$ (obtained with \texttt{ehtim} without any regularizer) as a starting point and Gaussian prior.
\end{itemize}
Although in this main text we show the Scenario C, the remaining scenarios can be found in the Appendix~\ref{app:m87_stponits}. We remark the importance of the initial points in real data, in particular with a sparse uv-coverage.

To avoid unnecessary extension of the paper, we just show the Pareto front and clustered images. The convergence of the algorithm has been already shown in Sec.~\ref{sec:synthetic_data}.

M87 was observed in four days April 6, 7, 10 and 11 (see for example Paper I). The reconstructions for all days can be found in Appendix~\ref{app:m87_fullepochs}.  

\begin{figure}
    \centering
    \hspace{-1cm}
    \includegraphics[width=0.5\textwidth]{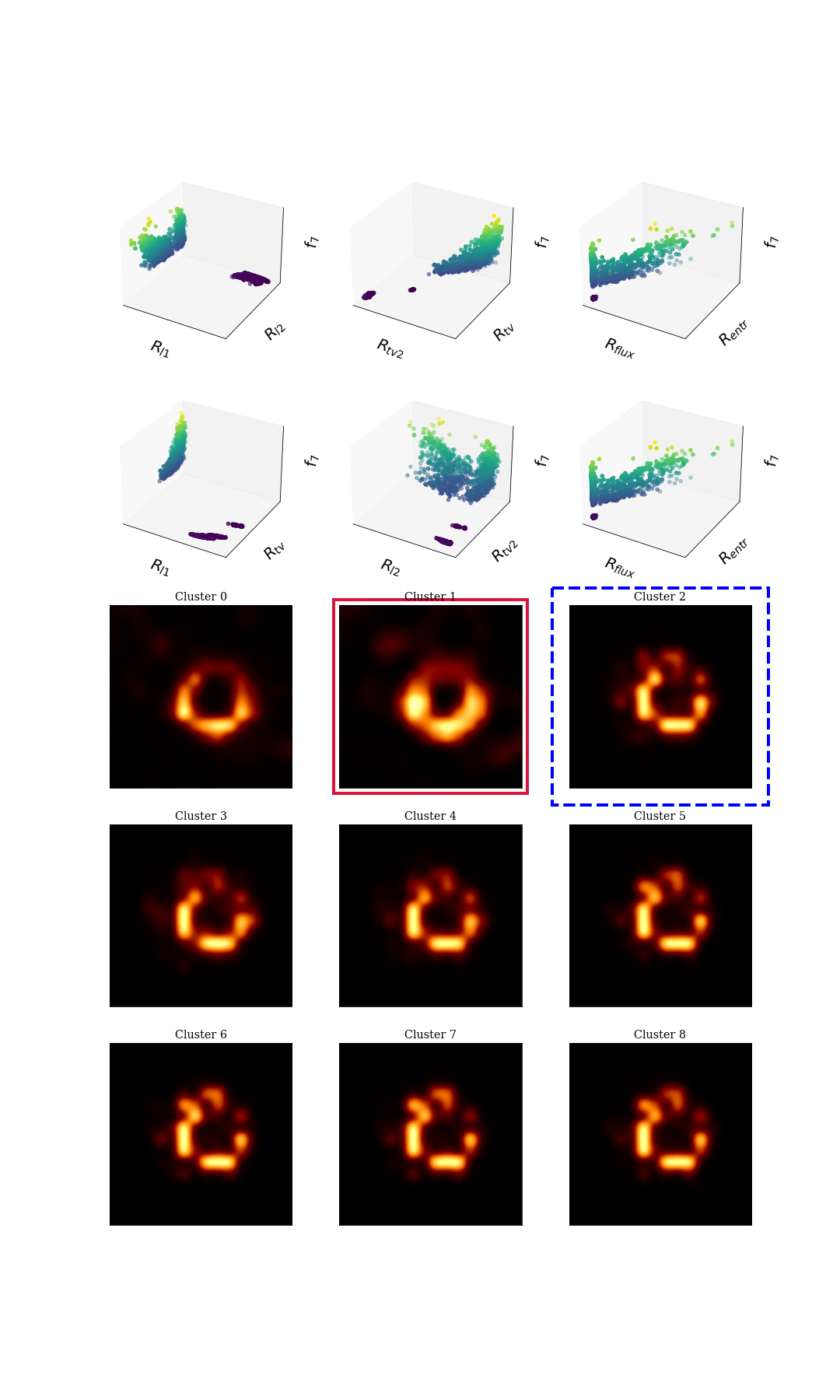}
    \vspace{-2cm}
    \hspace{-1.15cm}
    \caption{Relative Pareto fronts (top two lines) and clustered images for M87 in April 11 (bottom three lines).}
    \label{fig:m87_comp_rect}
\end{figure}

Figure~\ref{fig:m87_comp_rect} depicts the set of clustered solutions for the day April 11. %Fig.~\ref{fig:synth_uvcov} left panel shows the uv-coverage for this day. 
Cluster 0 and cluster 1 present a good ring structure. Indeed, the representant for cluster 1 is chosen as a preferred image reconstruction, which is very similar to the one published by the EHT. The rest of clusters have more subtle differences. We can observe this three ``families'' in the Pareto front.

\section{Closure-only Imaging}
\label{sec:closure_only}

During this paper we have considered $\alpha$ to be zero, and $\beta,\gamma,\delta$ being different from zero. Hence, the reconstruction is independent from the highly unstable phase calibration, but is only properly working if the dataset is amplitude self-calibrated as described in~\cite{Readhead1978}. Self-calibration aims to adjust the complex gains of an interferometer, by iteratively comparing the calibrated visibilities to an improved model. An accurate modeling of the source structure is crucial for the correct convergence of self-calibration~\citep[see for instance][]{Marti2008, mus22}.

In this section we try our algorithm in non-selfcalibrated data for the April 11 real EHT data and EHT + ngEHT data. For this case, when not self-calibration is applied, the non-closure related fitting weights ($\alpha$ and $\beta$) are set to zero, i.e. only the closure phases and the closure amplitudes are fitted as suggested by \citet{Chael2018, Mueller2022}. Moreover, setting $\alpha=\beta=0$ solves the redundancy of data terms that we mentioned in Sec. \ref{sec:vlbibases}.

The reconstruction problem becomes more challenging since the number of independent closure phases/closure amplitudes is smaller than the number of independent visibilities, i.e. the data consistency is less constraining \citep{Kulkarni1991, Chael2018, Blackburn2020}. Moreover, we like to mention that $S_{\text{cph}}$ and $S_{\text{cla}}$ are only approximations to the true likelihoods \citep{Lockhart2022}.

\begin{figure}
    \centering
    \includegraphics[width=0.5\textwidth]{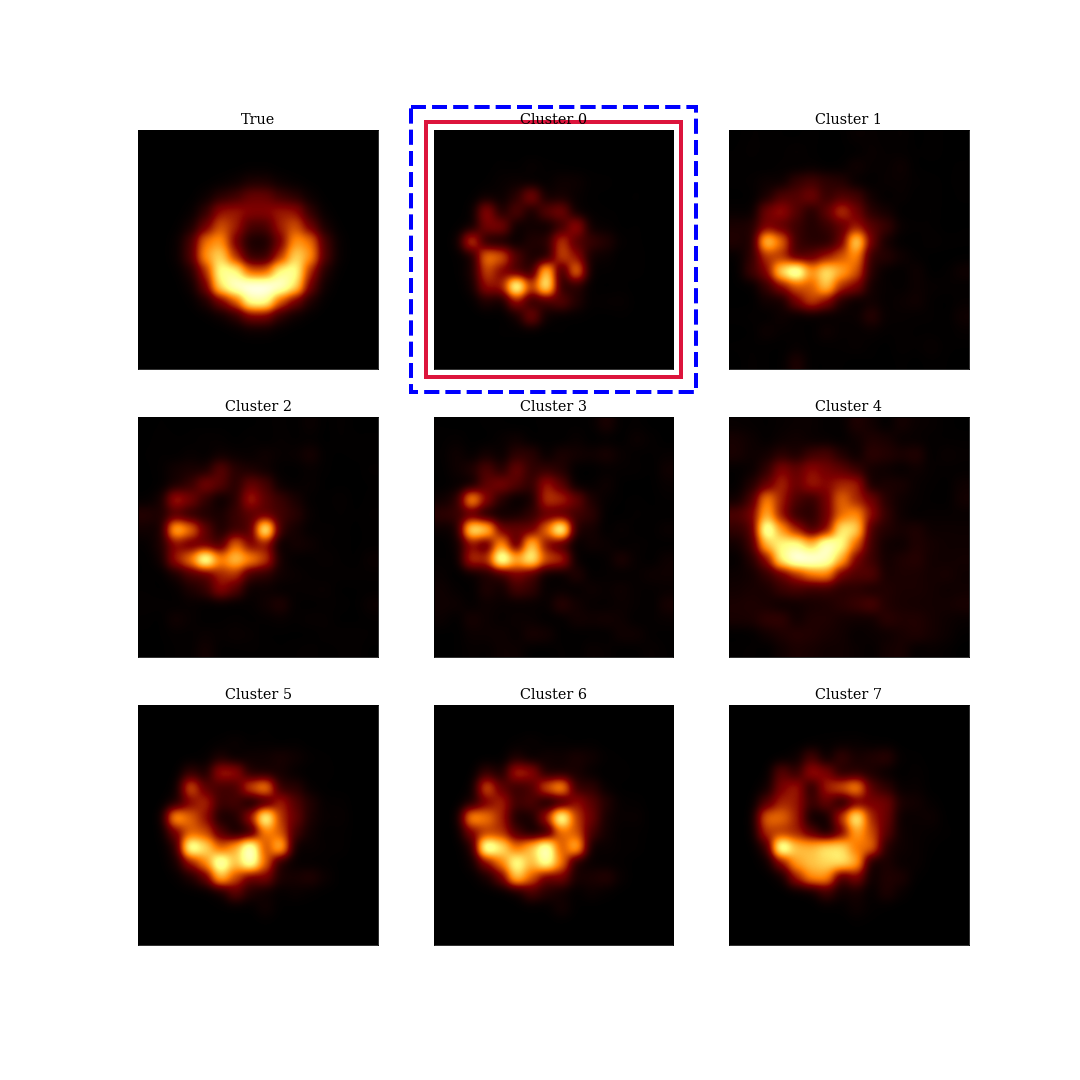}
    \vspace{-1cm}
    \caption{Cluster solutions for the case of the EHT+ngEHT data with noise included and not self-calibrated. The starting point is a random distribution in the pixels.}
    \label{fig:closure_only}
\end{figure}

\begin{figure*}
    \centering
    \includegraphics[width=0.4\textwidth]{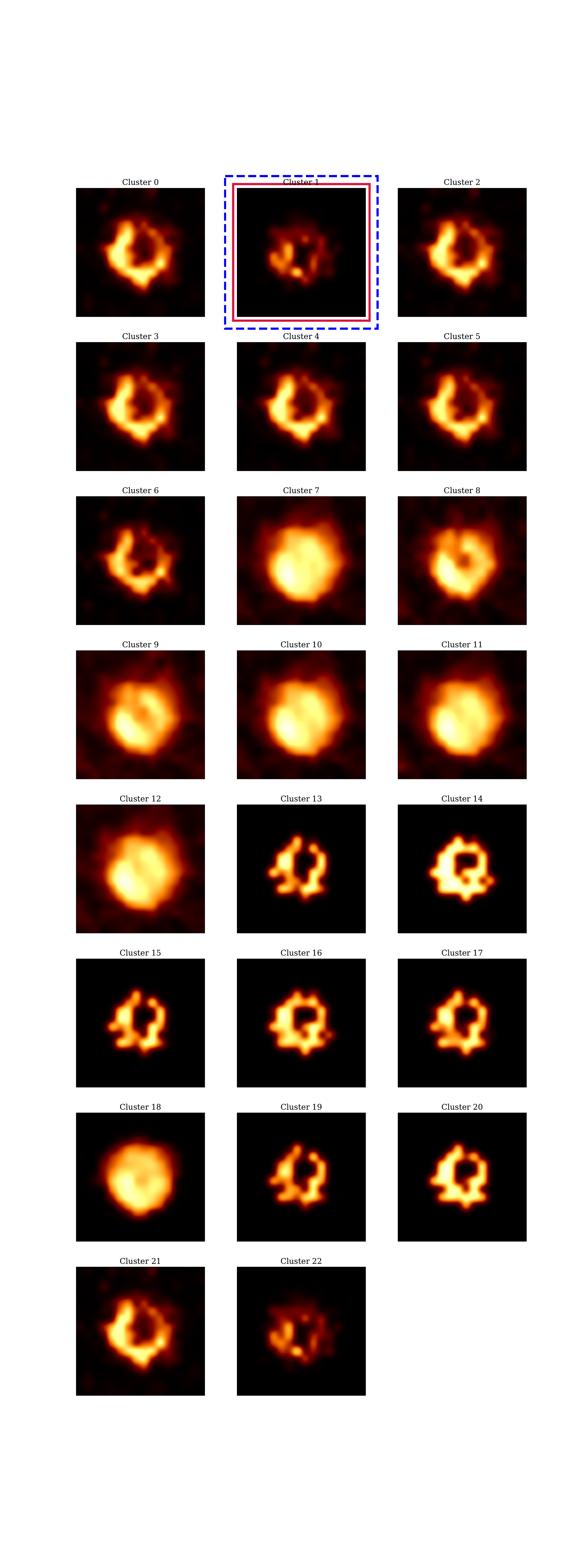}
    \includegraphics[width=0.4\textwidth]{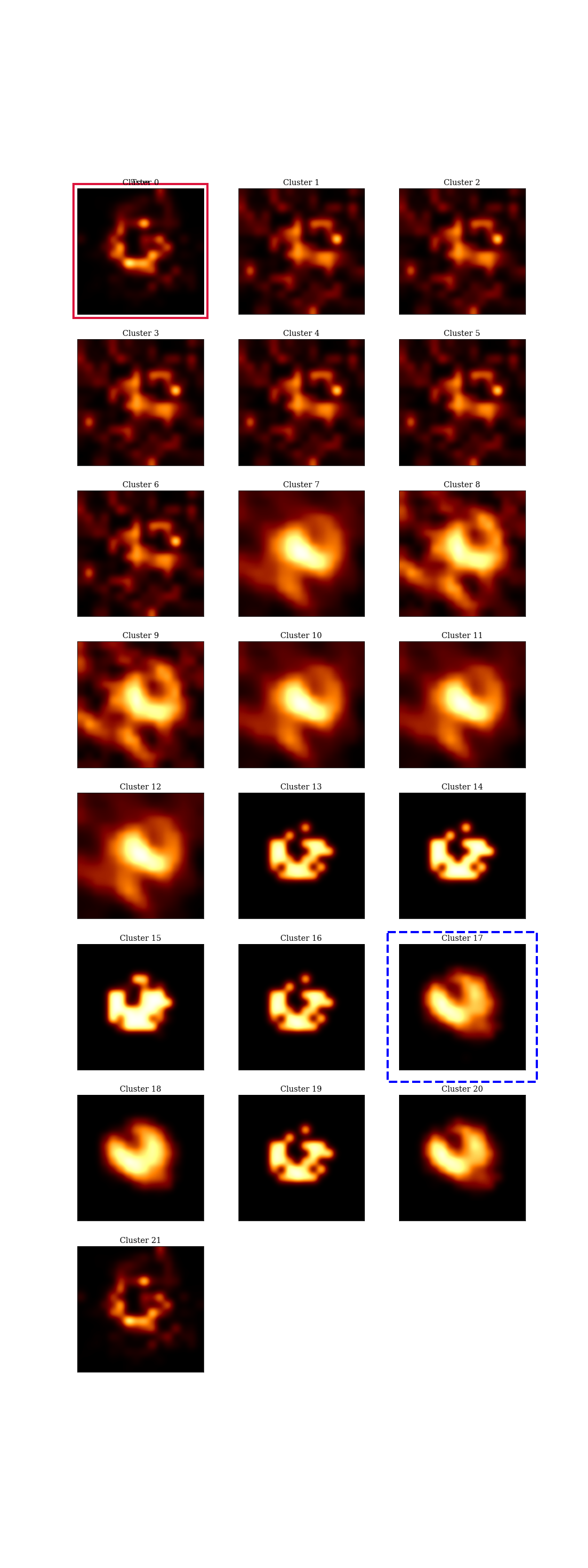}
    \vspace{-2cm}
    \caption{Clusters of solutions for the case of M87 in April 11 2017 EHT campaign. The starting points are a ring (left panel) and Gaussian (right panel) and play an important role to constrain the solutions in absence of a self-calibration model.}
    \label{fig:closure_only_real}
\end{figure*}

Figure~\ref{fig:closure_only} shows the set of solutions recovered using the EHT + ngEHT array using a random brightness distribution on the pixels as starting point. We observe that MOEA/D is able to recover the intrinsic source structure even if the self-calibration is not performed, although the obtained images are not as good as the ones obtained with a self-calibrated amplitude data term (i.e. $\beta \neq 0$). We note that the selection criteria for optimal solutions seem to select not the optimal solution in this case.

In Fig.~\ref{fig:closure_only_real}, we present the reconstructions for the real M87 EHT data. The poorest the uv-coverage, the less constrained is the optimization problem. In the case of self-calibrated data, the problem is biased by the self-calibration model. Instead, if there is not a ``biased'' model, there are more degrees of freedom, due to the smaller number of independent closure quantities. That is translated in more local minima and, in consequence, there are more clusters. We like to highlight that the starting point is even more crucial in the case of non-self-calibrated data, i.e. without amplitude consistency, in particular when the uv-coverage is sparse. Since a closure-only data set is less constraining than fitting fully calibrated visibilities, it is harder for the method to converge by random mutation and random, genetic mixing. For the left panel of Fig.~\ref{fig:closure_only_real} we used a ring as the starting point  and for the right panel, a Gaussian distribution. While in the first case we recover a ring with a central depression in most of the clusters, in the second one less ring-like structure are recovered, but still, it can be seen. Hence, the intrinsic structure of the source is predominant in the data even considering as starting point a distribution not related with the real structure.

The quality of the obtained solutions is worst when only closure quantities are fitted. Nevertheless, we can use one of the clusters for creating a non-biased model for self-calibration and rerun MOEA/D with updated amplitude information. Another alternative, is to use this picked cluster as initial point and run the MOEA/D. In this way, the MOEA/D is run iteratively improving the starting point.

\section{Summary and Conclusions}
\label{sec:summary}

Imaging in radioastronomy is an ill-posed inverse problem, and it is in particular hard when the uv-covarage is very sparse, as happens in the case of global VLBI observations. Several strategies have been developed that can be classified in three main families: CLEAN methods, (constrained) nonlinear optimization methods, and Bayesian approaches. Each algorithm has its advantages and disadvantages. For example, optimization methods are considerably faster than Bayesian ones, but lack a global posterior exploration. Therefore, a large, highly costy parameter survey was required. On the other side, Bayesian methods, explore a huge set of parameters, but they have slow performance compared to RML methods and CLEAN, in particular for large data sets. We identify two specific issues related to the imaging problem: the problem is multimodal and multiobjective.

In this work, we have presented a novel multiobjective formulation based on evolutionary algorithms that overcomes these problems. We compute a set candidate solutions that are non-dominated, i.e. ``optimally-balanced'': the Pareto front. A parameter survey is not required anymore since the complete Pareto front of optimal solutions is evolved in parallel, speeding-up considerably the time required by standard RML approaches to get a set of solutions. The result of parameter surveys depends on the set of test images. This issue does not arise for MOEA/D.

Moreover, the genetic algorithm is a global search technique that is less likely to get trapped into local extrema. Therefore, we are able to recover a full subset of solutions that are the best compromise. Every candidate solution in the Pareto front is related to a specific hyper-parameter combination.

MOEA/D is faster than any Bayesian approach, but does not explore a posterior distribution. Moreover, it is very flexible, allowing to introduce new regularizers very easily without increasing the computing complexity exponentially. We have created a clustering algorithm to group the ``similar'' solutions. Then, we have implemented two different techniques to choose the representative image between all clusters. All of the cluster are mathematically valid images, and therefore any other criterion to choose one among all can be also used.

We have successfully tested our algorithm in four synthetic models (double, disk, ring and crescent) using a sparse array (EHT 2017) and a more complete one (EHT + ngEHT). Finally, we have run our algorithm in real 2017 EHT M87. In this work we discussed the role of various regularization terms and their treating in a multiobjective framework. In a consecutive work we will focus on the data terms, i.e. include a wider variety of (non-redundant) data term combinations in the multiobjective formulation (also including dynamic and polarimetric data products), study their role in more details, and extend MOEA/D to dynamic and polarimetric observations.

% \begin{itemize}
%     \item No survey needed
%     \item Set of local minima recovered
%     %\item Metaheuristic, then it does not depend on the initial points and other parameters. It is a generic way, without supposing anything (only the a priori distribution which is needed for the $\chi^2$) on the function
% \end{itemize}

\section*{Software Availability}
We will make our imaging pipeline and our software available soon in the second release of MrBeam \footnote{\url{https://github.com/hmuellergoe/mrbeam}}. Our software makes use of the publicly available ehtim \citep{Chael2016, Chael2018}, regpy \citep{regpy}, MrBeam \citep{Mueller2022, Mueller2022b, Mueller2022c} and pygmo \citep{Biscani2020} packages.

\begin{acknowledgement}
AM and HM have contributed equally to this work. This work was partially supported by the M2FINDERS project funded by the European Research Council (ERC) under the European Union’s Horizon 2020 Research and Innovation Programme (Grant Agreement No. 101018682) and by the MICINN Research Project PID2019-108995GB-C22. AM also thank the Generalitat Valenciana for funding, in the frame of the GenT Project CIDEGENT/2018/021 and the Max Plank Institut für Radioastronomie for covering the visit to Bonn which has made this possible. HM received financial support for this research from the International Max Planck Research School (IMPRS) for Astronomy and Astrophysics at the Universities of Bonn and Cologne. The authors thank Michael Janssen for his valuable comments to improve this work.
\end{acknowledgement}

\bibliography{lib}{}
\bibliographystyle{aa}

\appendix

\section{Parameter Survey} \label{sec: parameter_survey}
The genetic evolution from one generation to the next generation in MOEA/D is done by genetic mixing and random mutation. These operations are controlled by specific control parameters. For full details we refer to \citet{Li2009}. Following the algorithmic outline presented in Sec. \ref{sec: moead}. Assume that $k,l$ are randomly selected indices in the neighborhood $U_B(\lambda^j)$. We aim to compute a new solution $y^j$ by genetic mixing and random mutation. The genetic mixing is \citep{Li2009}:
\begin{align}
y^j_i = \begin{cases}
x^j_i + F \cdot \left( x^k_i-x^l_i \right) & \text{with probability CR} \\
x^j_i & \text{with probability 1-CR}
\end{cases}
\end{align}
Random mutation can be written as \citep{Li2009}:
\begin{align}
y^j_i = \begin{cases}
x^j_i + \sigma_i \cdot \left( b_i-a_i \right) & \text{with probability } p_m \\
x^j_i & \text{with probability } 1-p_m
\end{cases},
\end{align}
where $a_i$ and $b_i$ are the lower and upper bound of the current decision vector. The magnitude of mutation is controlled by $\sigma_i$ \citep{Li2009}:
\begin{align}
\sigma_i = \begin{cases}
(2 \cdot \text{rand})^{\frac{1}{\eta+1}}-1 & \text{if rand}<0.5 \\
1-(2-2 \cdot \text{rand})^{\frac{1}{\eta+1}} & \text{otherwise}
\end{cases},
\end{align}
where $rand$ is a random number uniformly distributed in $[0,1]$. So, all in all we have four control parameters, two related to the genetic mixing ($F, CR$) and two related to the polynomial mutation ($p_m, \eta$). Moreover, the number of genes per generation and the number of generations are free parameters. A false combination of hyper-parameters could keep back MOEA/D from convergence (e.g. if the random mutation is to small) or from diversity in the solution (e.g. if the genetic mating appears to often). Overall, we started from the default choices suggested in \citet{Li2009} and surveyed several parameter combination with the crescent example with EHT coverage. We changed $F \in [0.1, 0.5, 0.9]$, $\eta \in [5, 20, 50]$ and changed the grid of the weighting combinations to $10^7, 30^7, 100^7$ and $500^7$ weight combinations. Moreover, we tested $1000$, $3000$ and $5000$ iterations. Except for parameters far out of the standard range (i.e. $F=0.9$), the performance was overall quite similar, with a tendency towards more generations, a smaller number of weight arrays, and genetic parameters close to the one that were found optimal in \citet{Li2009}. So we used $CR=1$, $F=0.5$, $\eta=20$ and $p_m=0.9$ for this manuscript. The number of weight parameter combinations was limited to $10^7$ combinations to keep the algorithm reasonable fast. Moreover, our investigation of more parameter combinations did not suggest an improved performance. Using $3000-5000$ iterations seems sufficient as also backed by the frozen in convergence of the decision vectors at these generations.

In MOEA/D we solve problems of the form:
\begin{align}
x^j \in \mathrm{argmin}_{x} \sum_{i=1}^m \lambda_i^j f_i(x)\,,
\end{align}
with the objective functionals discussed in Sec. \ref{sec: modelization}. Every point in the Pareto front represents a solution optimal with respect to the local parameter combination. In this way we test the output of RML imaging with several parameter configurations. However, due to the normalization of $\lambda^j$ and the limited gridsize, the weighting parameters for different optimization terms differ only by one order of magnitude. Since the regularization terms are not normalized with respect to the pixel size (e.g. TV-pseudonorm or $l^1$-norm change when a smaller pixel size is used) we have to rescale the regularization terms to a similar order of magnitude of impact before running MOEA/D. We varied the prefactor for $R_{l^1}$, $R_{l^2}$, $R_{TV}$ and $R_{TSV}$ between $[1, 10, 100]$ and the prefactor for entropy regularization in $[0.1, 1, 10]$. The parameter combinations were tested on synthetic EHT and EHT + ngEHT data for all four models with optimal combinations found for entropy rescaling by $0.1$ or $1$, $l^2$ rescaling by a factor of $10$, $l^1$ rescaling by a factor of $1$ and total variation and total squared variation rescaling by a factor of $1$ or $10$.

\section{M87 in all epochs}
\label{app:m87_fullepochs}

In this Appendix we extend the results of M87 shown in Sect.~\ref{sec:real_data} with the reconstructed images for the rest of the epochs.

\begin{figure*}
    \centering
    \includegraphics[width=0.3\textwidth]{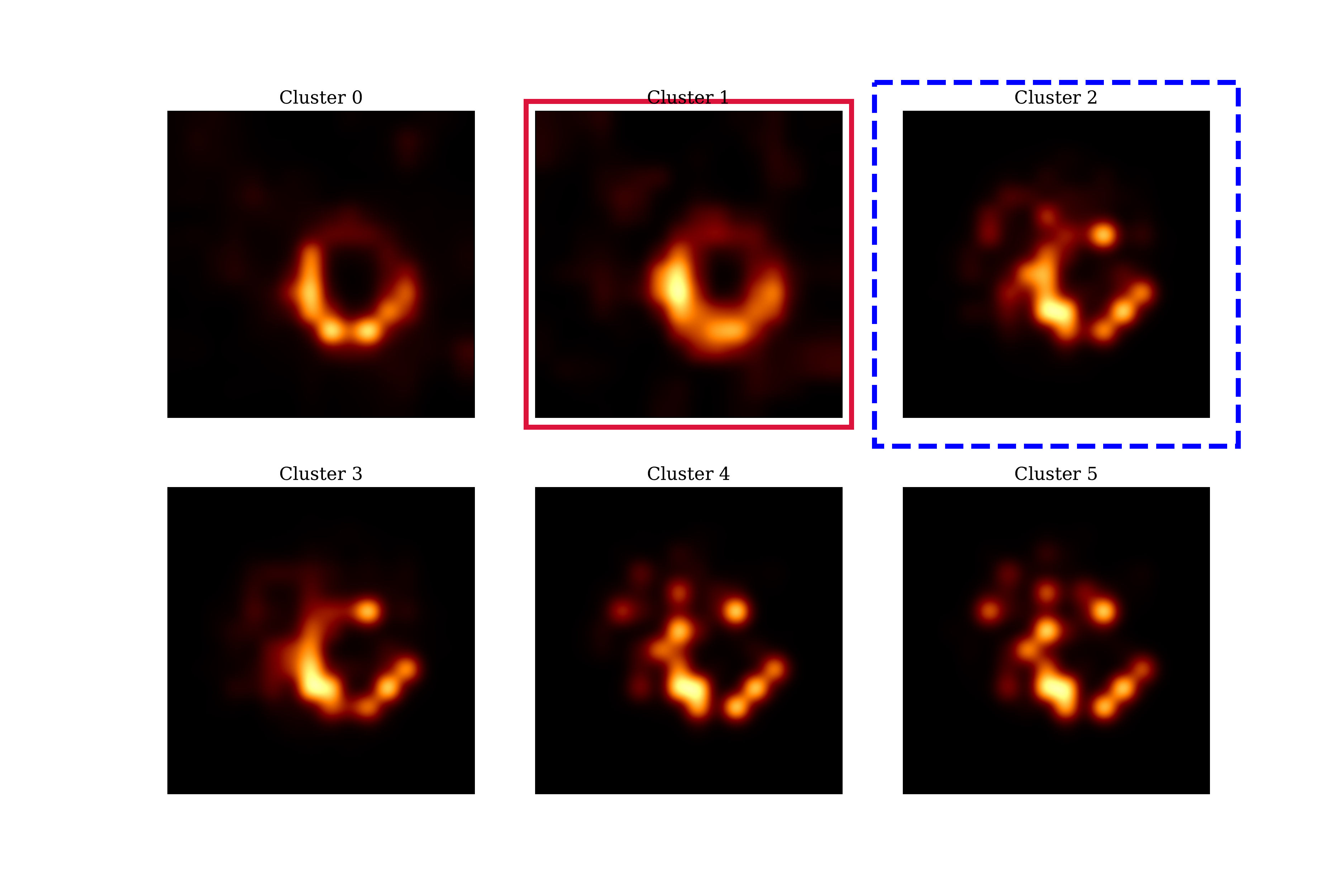}\\
    \includegraphics[width=0.3\textwidth]{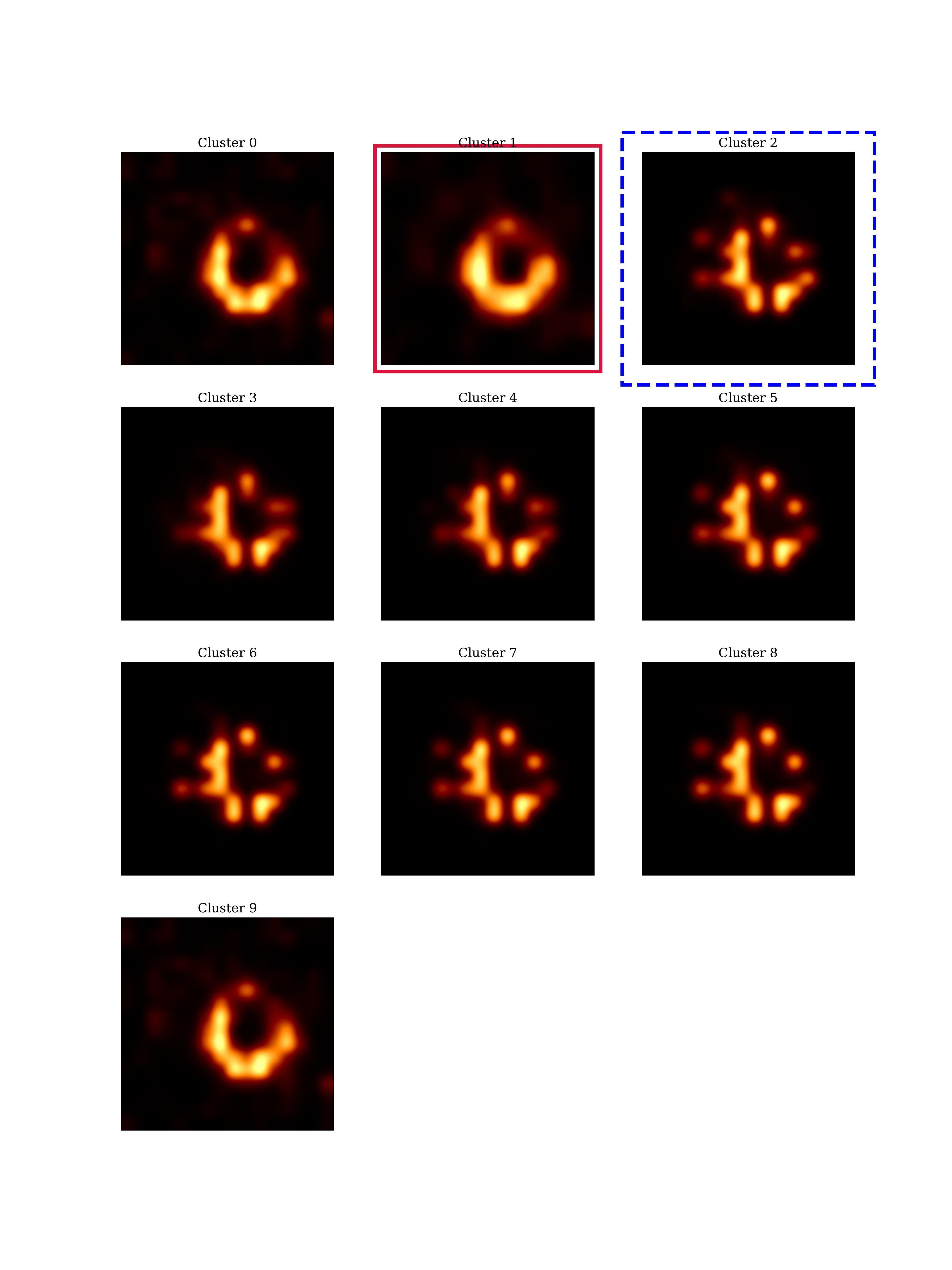}\\
    \includegraphics[width=0.3\textwidth]{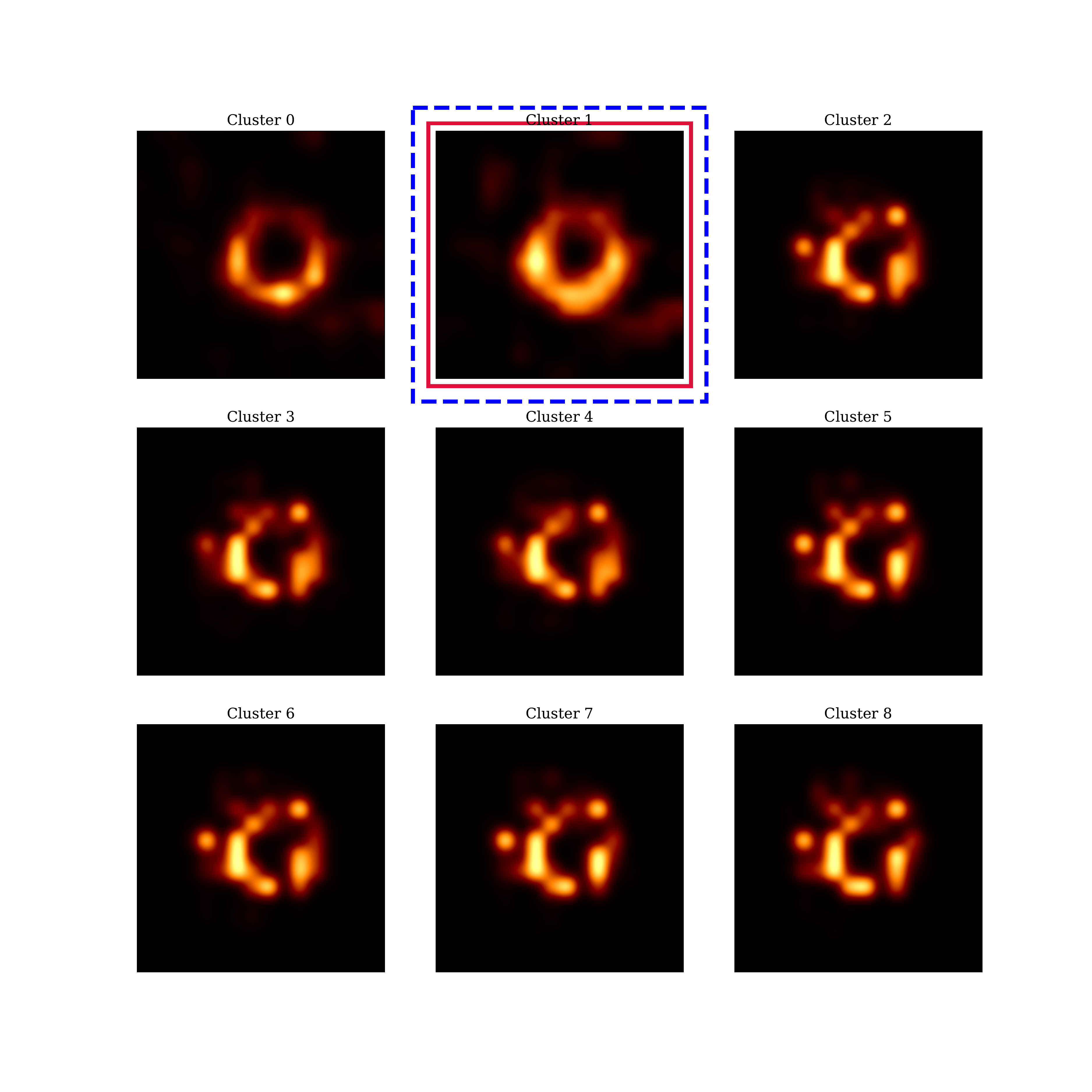}
    \caption{Image reconstructions of M87 2017 EHT observations. From top to bottom: April 5, April 6 and April 10.}
    \label{fig:m87_alldays}
\end{figure*}

Figure~\ref{fig:m87_alldays} depicts the image reconstructions for M87 in April 5, 6 and 10. In all of them the preferred image is the one presenting the clearest ring structure.

\section{M87 reconstructed using different starting points}
\label{app:m87_stponits}

In this Appendix we present two reconstructions for M87 April 11 using: 1) random brightness pixel distribution, 2) ring model as starting points

\begin{figure*}
    \centering
    \includegraphics[width=0.4\textwidth]{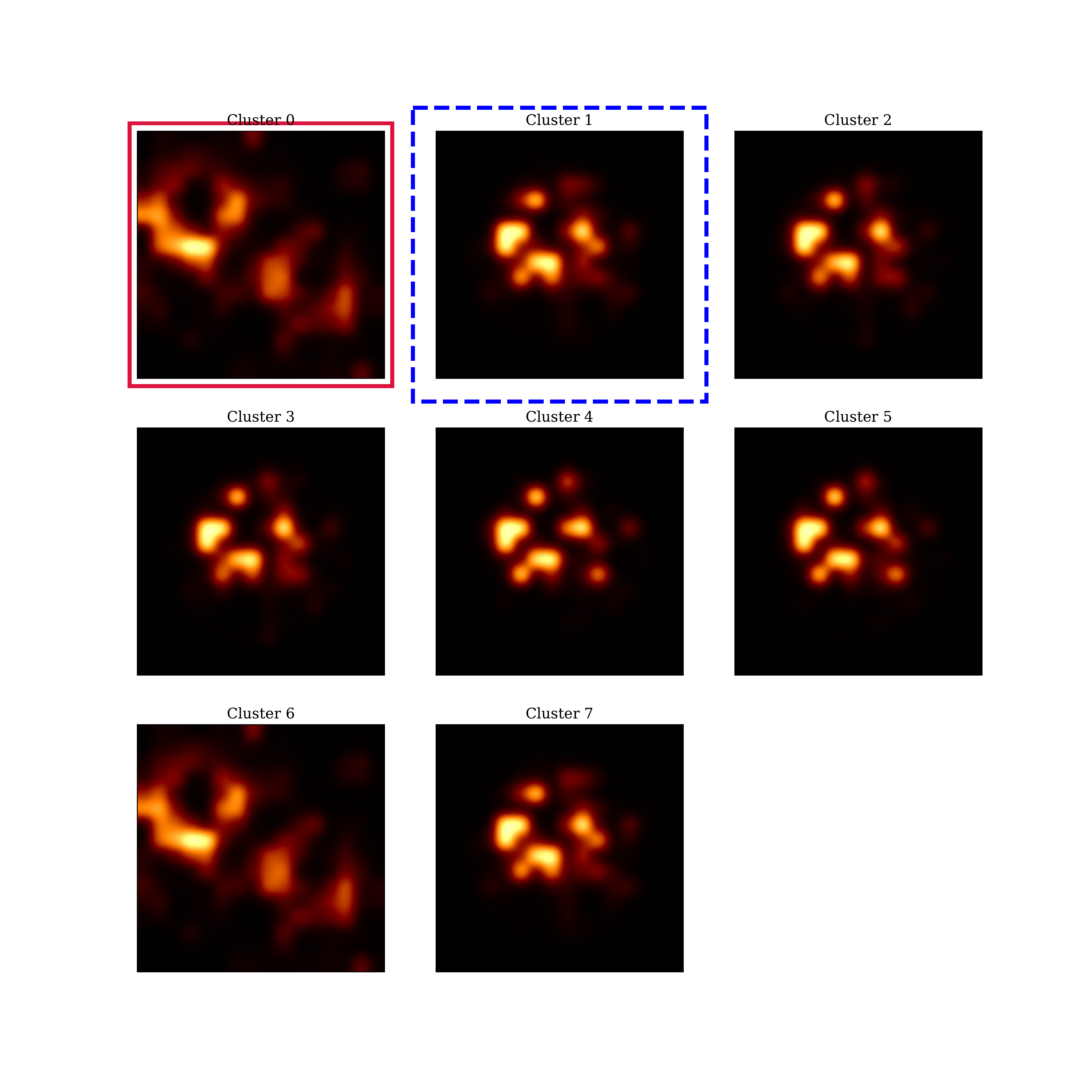}
    \includegraphics[width=0.4\textwidth]{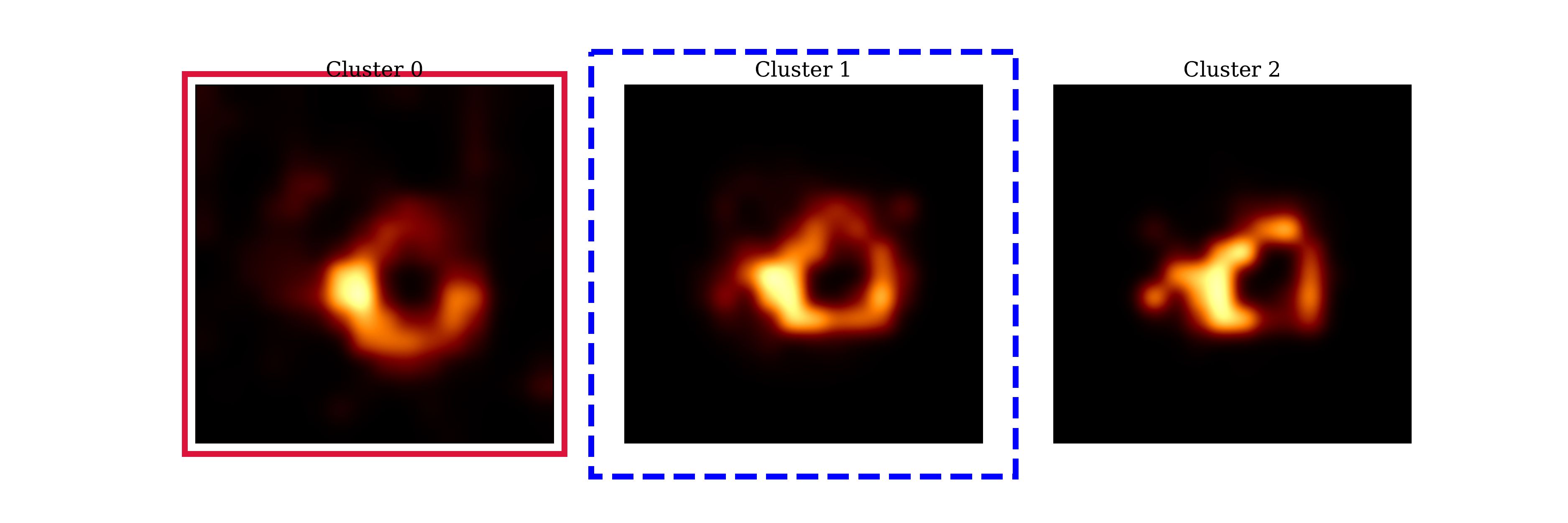}
    \caption{Set of solutions obtained supposing a random starting point (left panel) and ring starting point (right panel).}
    \label{fig:m87_stpoints}
\end{figure*}

Figure~\ref{fig:m87_stpoints} shows the different clusters of the obtained solutions. As expected, starting with a ring morphology, helps the convergence of algorithm. Nevertheless, even starting with a random brightness distribution on the pixels, we can recover a ring structure, which is a robust signature of the presence of a ring in the data. It is also remarkable that the less diversity presented with the geometric model. This could be due to the robust constraints that this initial starting point imposes.

\end{document}